\documentclass[
    reprint,
    aps,
    prl,
    superscriptaddress,
    amsmath,
    amssymb,
    floatfix,
    nofootinbib
]{revtex4-2}

\usepackage[T1]{fontenc}
\usepackage[utf8]{inputenc}
\usepackage{lmodern}
\usepackage{graphicx}

\usepackage[justification=raggedright,singlelinecheck=false]{caption}
\usepackage{subcaption}
\usepackage{dcolumn}
\usepackage{booktabs}
\usepackage{array}
\usepackage{multirow}
\usepackage{bm}
\usepackage{xcolor}
\colorlet{revision}{black} 
\colorlet{REVISION}{revision} 
\usepackage{comment}
\usepackage{slashed}

\usepackage[hyperfootnotes=false]{hyperref}

\graphicspath{{./}}

\newcommand{\bnl}{Physics Department, Brookhaven National Laboratory, Upton, NY 11973, USA}
\newcommand{\cu}{Physics Department, Columbia University, New York, NY 10027, USA}
\newcommand{\riken}{RIKEN-BNL Research Center, Brookhaven National Laboratory, Upton, NY 11973, USA}
\newcommand{\edinb}{School of Physics and Astronomy, The University of Edinburgh, Edinburgh EH9 3FD, UK}
\newcommand{\uconn}{Physics Department, University of Connecticut, Storrs, CT 06269-3046, USA}
\newcommand{\soton}{School of Physics and Astronomy, University of Southampton, Southampton SO17 1BJ, UK}
\newcommand{\innovation}{Collaborative Innovation Center of Quantum Matter, Beijing 100871, China}
\newcommand{\chep}{Center for High Energy Physics, Peking University, Beijing 100871, China}
\newcommand{\pkuphy}{School of Physics, Peking University, Beijing 100871, China}
\newcommand{\scnt}{Southern Center for Nuclear-Science Theory (SCNT), Institute of Modern Physics, Chinese Academy of Sciences, Guangdong 516000, China}

\hypersetup{
    colorlinks=true,
    linkcolor=blue,
    citecolor=blue,
    urlcolor=blue,
    pdftitle={First Lattice Calculation of Structure-Dependent Radiative Corrections to Lepton-Flavor-Universality Ratios in Leptonic Decays},
    pdfauthor={Peter Boyle, Norman H. Christ, Xu Feng, Taku Izubuchi, Luchang Jin, Christopher T. Sachrajda, Xin-Yu Tuo},
    pdfsubject={Lattice QCD radiative corrections}
}

\begin{document}

\preprint{}

\title{\textcolor{revision}{First Lattice Calculation of Structure-Dependent Radiative Corrections to Lepton-Flavor-Universality Ratios in Leptonic Decays}}

\author{Peter Boyle}
\affiliation{\bnl}
\affiliation{\edinb}

\author{Norman H. Christ}
\affiliation{\cu}

\author{Xu Feng}
\affiliation{\pkuphy}
\affiliation{\innovation}
\affiliation{\chep}
\affiliation{\scnt}

\author{Taku Izubuchi}
\affiliation{\bnl}
\affiliation{\riken}

\author{Luchang Jin}
\affiliation{\uconn}

\author{Christopher T. Sachrajda}
\affiliation{\soton}

\author{Xin-Yu Tuo}
\email{ttxxyy.tuo@gmail.com}
\affiliation{\bnl}

\date{November 13, 2026}

\begin{abstract}
The ratio of electronic to muonic leptonic decay widths, $R_{e/\mu}$, for the light mesons $\pi$ and $K$, provides a clean test of lepton flavor universality (LFU) and a sensitive probe of physics beyond the Standard Model. Its Standard-Model prediction is exceptionally precise, with the leading uncertainty associated with the structure-dependent (SD) radiative correction of $O(0.1\%)$. As experiments such as PIONEER and NA62 aim for unprecedented precision, this SD correction has become an essential ingredient in precision experiment--theory comparisons. We present the first lattice QCD$+$QED calculation of this SD correction at the physical pion mass and in the continuum limit. We employ the infinite-volume reconstruction (IVR) method with Coulomb-gauge photons, significantly reducing both statistical errors and finite-volume effects.
We obtain the Standard-Model predictions,
$R_{e/\mu}=\textcolor{revision}{1.23516(10)}\times10^{-4}$ for $\pi$ and
$R_{e/\mu}=\textcolor{revision}{2.47661(31)}\times10^{-5}$ for $K$. Our results reduce the hadronic uncertainty in $R_{e/\mu}$, provide the most precise Standard-Model predictions to date, and establish first-principles benchmarks for future high-precision tests of LFU.
\end{abstract}

\maketitle

\par\medskip
\noindent\textbf{Introduction -}
Precision tests of lepton flavor universality (LFU), the universality of
the charged-lepton gauge couplings in the Standard Model, provide a
powerful indirect probe of physics beyond the Standard Model~\cite{Bryman:2021teu}. A particularly sensitive LFU test is the ratio of the electronic and
muonic leptonic decay widths,
\begin{equation}\label{Remu_def1}
      R_{e/\mu} \equiv \frac{\Gamma\left(P \to
  e\bar{\nu}_e(\gamma)\right)}{\Gamma\left(P \to \mu\bar{\nu}_\mu(\gamma)\right)},\quad
  P=\pi,\,K.
  \end{equation}
Hereafter, the $P$ dependence of $R_{e/\mu}$ is left implicit; for $P=K$, we follow the convention of Ref.~\cite{Cirigliano:2007xi} and exclude the structure-dependent (SD) real-photon emission from $R_{e/\mu}$.
The helicity suppression of the electronic channel, reflected in the smallness of $R_{e/\mu}$, historically provided key evidence for the $V\!-\!A$ structure of the weak interaction~\cite{Impeduglia:1958gha,Fazzini:1958ii,Feynman:1958ty,Sudarshan:1958vf,Fidecaro:2015pvi}. This suppression makes
$R_{e/\mu}$ exceptionally sensitive to new physics: scalar or
pseudoscalar interactions beyond the Standard Model are not helicity
suppressed, so their relative effects can be enhanced, allowing
$R_{e/\mu}$ to probe mass scales beyond the direct reach of
colliders~\cite{Shanker:1982nd,Campbell:2003ir,Campbell:2008um,Cirigliano:2012ab}. 

Another key advantage of $R_{e/\mu}$ in LFU tests is its exceptionally clean theoretical prediction. The dominant
uncertainty comes from the SD part of the
$O(\alpha)$ radiative corrections, whose contribution is suppressed to
the $O(0.1\%)$ level by the cancellation of hadronic effects between the
$e$ and $\mu$ channels. This correction has been computed in chiral
perturbation theory (ChPT), leading to Standard-Model predictions for
$R_{e/\mu}$ with an accuracy of $0.01\%$~\cite{Cirigliano:2007xi,Cirigliano:2007ga}. 

These advantages have motivated increasingly precise measurements of
$R_{e/\mu}$. The most precise measurements to date come from PIENU at TRIUMF
for the pion~\cite{PiENu:2015seu} and NA62 at CERN for the kaon~\cite{NA62:2012lny}. Nevertheless, current measurements remain approximately an order of magnitude less precise than the theoretical prediction, motivating the next generation of experiments:
PIONEER aims to reach a precision of $0.01\%$ in the pion channel~\cite{PIONEER:2022yag,PIONEER:2022alm,PIONEER:2025idw},
comparable to the current theoretical precision, while NA62 is pursuing an updated measurement of the kaon ratio with a projected precision of about $0.2\%$~\cite{Massri:2025NA62}.

Although the SD correction enters only at the $0.1\%$ level, it is
non-negligible at the precision targeted by future experiments. Its
calculation is challenging because it requires non-perturbative QCD
input. In the ChPT prediction~\cite{Cirigliano:2007xi,Cirigliano:2007ga}, the dominant uncertainty arises from truncating the chiral expansion. For example, at the order considered, the form factors $F_V$ and $F_A$ governing the SD real-photon
emission are constant, with photon-momentum
dependence entering only at higher orders. Lattice QCD$+$QED enables a
first-principles determination of this correction, while providing an independent check of the ChPT prediction; such a determination becomes increasingly important as future experimental precision
approaches that of the Standard-Model prediction.

Previous lattice-QCD studies of radiative corrections to light-meson
leptonic decays have focused mainly on the muon channels and the related
$K_{\mu 2}/\pi_{\mu 2}$ ratio~\cite{Carrasco:2015xwa,Giusti:2017dwk,DiCarlo:2019thl,Boyle:2022lsi}, as well as on the radiative
decays $P\to\ell\nu\gamma^{(*)}$~\cite{Desiderio:2020oej,Frezzotti:2020bfa,Gagliardi:2022szw,DiPalma:2025iud,Tuo:2021ewr,Boyle:2025uuh,DiPalma:2026tvl,DiPalma:2026fni}. The infinite-volume reconstruction (IVR) method removes the
power-law finite-volume effects associated with the massless photon,
changing their scaling to exponential suppression~\cite{Feng:2018qpx,Christ:2023lcc}. 
Our application of IVR to the isospin-breaking correction to the $K_{\mu 2}/\pi_{\mu 2}$ ratio will be reported in a separate publication. 
In the present work, we target the SD correction to $R_{e/\mu}$, which is about an order of magnitude smaller than the isospin-breaking correction to $K_{\mu 2}/\pi_{\mu 2}$. Consequently, the finite-volume and statistical errors are substantially magnified relative to the signal.

To further reduce these errors, we extend the IVR method with Coulomb-gauge photons. As explained below, this choice suppresses the point-like contribution, thereby reducing both statistical uncertainties and finite-volume effects.
We present the first lattice QCD determination of the SD radiative
contribution to $R_{e/\mu}$ at the physical
pion mass and in the continuum limit. 

\par\medskip
\noindent\textbf{Structure-dependent correction -}
\label{sec:def_challenge}
The calculation for $R_{e/\mu}$ can be organized in a
perturbative QED expansion~\cite{Cirigliano:2007xi,Cirigliano:2007ga},
\begin{equation}\label{Remu_def2}
\begin{aligned}
    R_{e/\mu} &= R_{e/\mu}^{(0)}\left[1+\Delta_{\alpha,\text{pt}}+\Delta_{\alpha,\text{SD}}+\Delta_{\alpha^{n\geq 2}}\right],\\
    R_{e/\mu}^{(0)} &= \frac{m_e^2}{m_\mu^2}\left(\frac{m_P^2-m_e^2}{m_P^2-m_\mu^2}\right)^2 .
\end{aligned}
\end{equation}
Here $R_{e/\mu}^{(0)}$ is the result in the absence of
electromagnetic corrections, $\Delta_{\alpha,\text{pt}}$ denotes
the $O(\alpha)$ correction in the point-like approximation whose form is known analytically~\cite{Marciano:1993sh}, and
$\Delta_{\alpha,\text{SD}}$ is the $O(\alpha)$ SD correction beyond the point-like approximation. Higher-order QED effects are collected in
$\Delta_{\alpha^{n\geq 2}}$; its leading-logarithmic (LL) result is given in Ref.~\cite{Marciano:1993sh}.

The leading theoretical uncertainty comes from
$\Delta_{\alpha,\mathrm{SD}}$, which is the target of this work.
We decompose it into virtual- and real-photon corrections,
\begin{equation}
\Delta_{\alpha,\mathrm{SD}}
=
\delta_{\mathrm{SD}}^{\mathrm{vir}}
+
\delta_{\mathrm{SD}}^{\mathrm{real}}\,.
\end{equation}
Throughout, the correction without explicit lepton-mass arguments denotes the electron-muon difference. In particular, $\delta_{\mathrm{SD}}^{X}$ denotes $ \delta^X_{\mathrm{SD}}(m_P,m_e) -\delta^X_{\mathrm{SD}}(m_P,m_\mu)$, where $X\in\{\mathrm{vir},\mathrm{real}\}$ and
$\delta^X_{\mathrm{SD}}(m_P,m_\ell)$ is the corresponding single-channel $O(\alpha)$ SD correction. 

\begin{figure}
    \includegraphics[width=\columnwidth]{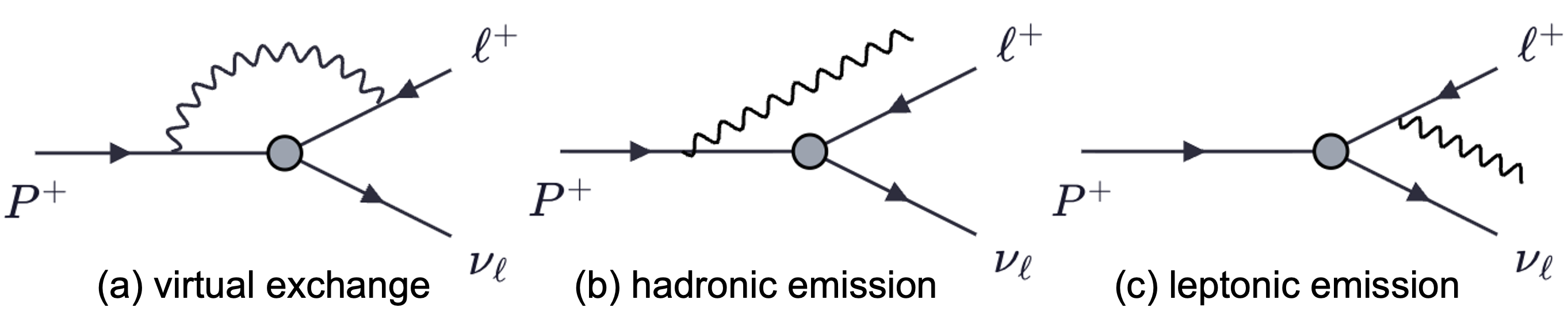}
    \caption{\label{fig:diagrams} Feynman diagrams relevant to the $O(\alpha)$ SD correction: (a) virtual-photon exchange between the initial-state meson and the final-state lepton; (b) real-photon emission from the meson; and (c) real-photon emission from the lepton.}
\end{figure}

The real-photon-emission correction
$\delta_{\mathrm{SD}}^{\mathrm{real}}$ consists of the
SD part of the squared amplitude corresponding to
Fig.~\ref{fig:diagrams}(b), together with the interference between the SD part of the amplitude in
Fig.~\ref{fig:diagrams}(b) and the amplitude in
Fig.~\ref{fig:diagrams}(c). The squared amplitude corresponding to 
Fig.~\ref{fig:diagrams}(c) is included in
$\Delta_{\alpha,\mathrm{pt}}$.
The correction $\delta_{\text{SD}}^{\text{real}}$ is
determined by the vector and axial-vector form factors, $F_V$ and $F_A$. These form factors are \textcolor{revision}{computed using the method of our previous work}~\cite{Boyle:2025uuh}; here we use them to perform the phase-space integration to obtain $\delta_{\text{SD}}^{\text{real}}$. The explicit
integrand and phase-space conventions are given in Sec.~\ref{app:real_photon} of the Supplemental Material~\cite{SupplementalMaterial}. 

The virtual correction $\delta_{\text{SD}}^{\text{vir}}$
arises from the interference between the tree-level amplitude and the
loop diagram shown in Fig.~\ref{fig:diagrams}(a). Apart from this diagram, the virtual correction to the external meson
leg is lepton-flavor independent and cancels in the $e-\mu$ difference, whereas the virtual correction to the external lepton
leg contains no SD contribution
and is included in $\Delta_{\alpha,\mathrm{pt}}$. 
The correction $\delta_{\text{SD}}^{\text{vir}}$ is given by
\begin{equation}\label{eq:delta_vir_def}
\begin{aligned}
    \delta_{\text{SD}}^{\text{vir}}(m_P,m_\ell)&=\delta^{\text{vir}}(m_P,m_\ell)-\delta_{\text{pt}}^{\text{vir}}(m_P,m_\ell),\\
    \delta^{\text{vir}}(m_P,m_\ell)&=\frac{2\pi\alpha}{f_P m_P^2 m_\ell^2(1-r_\ell)}\\
    &\times\int\frac{d^4k}{(2\pi)^4}\frac{L_{\mu}^{\sigma}(k,p,p_\ell)\,S_{\sigma\rho}(k)\, H_M^{\mu\rho}(k,p)}{(p_\ell-k)^2-m_\ell^2+i\epsilon},
\end{aligned}
\end{equation}
where $k$, $p$, and $p_\ell$ denote the photon, meson, and lepton momenta, respectively, and
$r_\ell=m_\ell^2/m_P^2$. Here $S_{\sigma\rho}(k)$ is the photon
propagator, and $L_\mu^{\sigma}(k,p,p_\ell)$ is the Dirac trace obtained after summing over the spins of the final-state leptons. The expressions are defined in Minkowski space. The subtraction term $\delta_{\text{pt}}^{\text{vir}}(m_P,m_\ell)$ is
given by replacing the Minkowski hadronic tensor
$H_M^{\mu\rho}(k,p)$ by its point-like approximation
$H_{M,\text{pt}}^{\mu\rho}(k,p)$. They are
defined as
\begin{eqnarray}
    &&H_M^{\mu\rho}(k,p)=\int\! d^4x\,e^{ik\cdot x}\langle 0|T\{J^{\text{EM},\rho}(x)J^{W,\mu}(0)\}|P(\vec 0)\rangle,\nonumber\\
    &&H_{M,\text{pt}}^{\mu\rho}(k,p)=f_P\Big[g^{\mu\rho}+\frac{(2p-k)^\rho(p-k)^\mu}{2p\cdot k-k^2}\Big],
    \label{eq:HM_def}
\end{eqnarray}
where the electromagnetic current is
$J^{\text{EM},\rho}=\tfrac{2}{3}\bar u\gamma^\rho u-\tfrac{1}{3}\bar d\gamma^\rho d-\tfrac{1}{3}\bar s\gamma^\rho s$,
and the weak current is
$J^{W,\mu}=\bar q\,'\gamma^\mu(1-\gamma_5)u$, with $q'=d$ for
$P=\pi$ and $q'=s$ for $P=K$.

The point-particle subtraction in Eq.~\eqref{eq:delta_vir_def} involves a large cancellation:
$\delta^{\text{vir}}(m_P,m_\ell)$ is dominated by the point-like contribution $\delta_{\text{pt}}^{\text{vir}}(m_P,m_\ell)$, leaving $\delta_{\text{SD}}^{\text{vir}}(m_P,m_\ell)$ only as a small remainder. As a result, statistical uncertainties and finite-volume effects are amplified relative to the SD signal. In the next section, we introduce a
method to suppress the point-like contribution and mitigate this
problem.

\par\medskip
\noindent\textbf{Coulomb-gauge IVR -}
\label{sec:virtual_photon}
We mitigate the amplification of uncertainties by using the IVR method with Coulomb-gauge photons. The key observation is that, by construction, the point-like contribution $H_{M,\text{pt}}^{\mu\rho}(k,p)$ satisfies the same Ward identity as the full hadronic matrix element,
$k_\rho H_M^{\mu\rho}(k,p)=k_\rho H_{M,\text{pt}}^{\mu\rho}(k,p)=f_P\,p^\mu$.
Their difference, namely the SD part, therefore obeys
$k_\rho\big(H_M^{\mu\rho}(k,p)-H^{\mu\rho}_{M,\text{pt}}(k,p)\big)=0$.
Consequently, when contracted with the photon propagator, the gauge-dependent term ($\propto k_\sigma k_\rho$) vanishes, and $\delta_{\text{SD}}^{\text{vir}}(m_P,m_\ell)$ is therefore gauge invariant. 
This gauge invariance allows us to choose a photon gauge in which the point-like contribution in Fig.~\ref{fig:diagrams}(a) is suppressed, thereby reducing the corresponding statistical errors and finite-volume effects. In this work, we use the Coulomb gauge,
\begin{equation}\label{eq:Scoul}
    S^{\text{Coul}}_{\sigma\rho}(k)=\frac{i}{\vec k^2}n_\sigma n_\rho+\frac{i}{k^2+i\epsilon}\sum_{\lambda=1,2}\varepsilon_\sigma(k,\lambda)\varepsilon_\rho^*(k,\lambda),
\end{equation}
where $n=(1,\vec 0)$, and $\varepsilon_\sigma(k,\lambda)$ with $\lambda=1,2$ are transverse polarization vectors satisfying $\varepsilon_0(k,\lambda)=0$, $\vec{k}\cdot \vec{\varepsilon}(k,\lambda)=0$, and $\vec{\varepsilon}(k,\lambda)\cdot \vec{\varepsilon}^{\,*}(k,\lambda')=\delta_{\lambda\lambda'}$.
The first and second terms on the right-hand side are the Coulomb-potential and transverse-photon contributions, respectively. 

The advantage of the Coulomb gauge is that the diagram shown in Fig.~\ref{fig:diagrams}(a) is free of infrared (IR) divergence, which strongly suppresses the point-like contribution. In the Feynman gauge, the IR divergence arises in the soft region of the loop integral,
$\int d^4k/[k^2(p\cdot k)(p_\ell\cdot k)]$. In the meson rest frame, however, Coulomb-gauge transverse photons satisfy
$\varepsilon(k,\lambda)\cdot k=\varepsilon(k,\lambda)\cdot p=0$, and hence
$\varepsilon_\rho^*(k,\lambda)H_{M,\text{pt}}^{\mu\rho}(k,p)=\varepsilon^{\mu*}(k,\lambda)f_P$.
The factor $(p\cdot k)$ is therefore absent from the denominator, making the transverse contribution IR finite. The Coulomb-potential propagator contains no $k^0$ pole; in the soft limit, the relevant meson and lepton poles approach the $k^0$ integration contour from the same side, so no pinch singularity develops and the Coulomb-potential contribution is also IR finite.%
\footnote{In Coulomb gauge, the remaining IR divergences arise only from the virtual correction to the external lepton leg and from real-photon emission off the final-state lepton (squared amplitude of Fig.~\ref{fig:diagrams}(c)). These parts contain no SD contribution.}

Moreover, although the two gauges agree in the continuum and infinite-volume limits, they are affected by different discretization and finite-volume effects due to the use of a local lattice electromagnetic current and the truncation of the infinite-volume photon propagator at the lattice boundary. Their agreement therefore provides a nontrivial consistency check between the two lattice implementations.

The IVR framework for radiative corrections to meson leptonic decays was developed in Feynman gauge in Ref.~\cite{Christ:2023lcc}.
Here we extend it to Coulomb gauge. In what follows, all quantities are defined in Euclidean space. After the Wick rotation, the Coulomb-gauge virtual correction can be written as
\begin{equation}\label{eq:delta_coul_lat}
\begin{aligned}
    \delta^{\text{vir}}_{\text{Coul}}(m_P,m_\ell)&=\int d^3x\!\int_{-t_s}^{\infty}\!dt\,H_{\mu\rho}(t,\vec{x})\,\mathcal{K}_{\mu\rho}(t,\vec x)\\
    &+\int d^3x~H_{\mu\rho}(-t_s,\vec{x}) \,\mathcal{K}^{(l)}_{\mu\rho}(-t_s,\vec x).
\end{aligned}
\end{equation}
\textcolor{revision}{Here, the time $-t_s$ separates the short-distance region, where the contribution is computed directly, from the long-distance region, where it is reconstructed from the lattice data at $t=-t_s$ using ground-state $P$-meson dominance.} \textcolor{revision}{Only the Coulomb-potential contribution requires reconstruction, since no single-meson intermediate state contributes to the transverse-photon part.} The functions $\mathcal{K}_{\mu\rho}(t,\vec x)$ and $\mathcal{K}^{(l)}_{\mu\rho}(-t_s,\vec x)$ are known Euclidean weight functions constructed from the leptonic trace and the lepton and Coulomb-gauge photon propagators. 
The Euclidean hadronic function is given by
\begin{equation}\label{IE}
    H_{\mu \rho}(t, \vec{x})  =\frac{\langle 0| T\{j_{\mu}^W(0) j_{\rho}^{\mathrm{EM}}(t, \vec{x})\}|P(\vec{0})\rangle}{\sum_{\vec{x}}\langle 0| j_{4}^W(0) j_{4}^{\mathrm{EM}}(-\Delta T, \vec{x})|P(\vec{0})\rangle}.
\end{equation}
Here $j_{\mu}^W$ and $j_{\rho}^{\mathrm{EM}}$ are Euclidean weak and electromagnetic currents. We normalize by the zero-momentum projection of the hadronic function at a separation $t=-\Delta T$, chosen large enough for ground-state dominance,
$\sum_{\vec{x}}\langle 0| j_{4}^W(0) j_{4}^{\mathrm{EM}}(-\Delta T, \vec{x})|P(\vec{0})\rangle=im_P\,f_P$. In practice, we evaluate the tensor convolution in Eq.~\eqref{eq:delta_coul_lat} using a scalar-function decomposition. The derivation of the resulting weight functions and the numerical evaluation details are given in Secs.~\ref{app:virtual_ivr} and \ref{app:weight_functions}. 

Subtracting the point-like contribution yields the SD correction
\begin{eqnarray}
    \delta_{\text{SD},\text{Coul}}^{\text{vir}}(m_P,m_\ell)&=&\delta_{\text{Coul}}^{\text{vir,pot}}(m_P,m_\ell)+\delta_{\text{Coul}}^{\text{vir,tr}}(m_P,m_\ell)\nonumber\\
    &&\hspace{0.3in}-\delta_{\text{pt},\text{Coul}}^{\text{vir}}(m_P,m_\ell),\label{eq:coul_SD}
\end{eqnarray}
where $\delta_{\text{Coul}}^{\text{vir,pot}}(m_P,m_\ell)$ and $\delta_{\text{Coul}}^{\text{vir,tr}}(m_P,m_\ell)$ denote the Coulomb-potential and the transverse-photon contributions, respectively. The point-like term $\delta_{\text{pt},\text{Coul}}^{\text{vir}}(m_P,m_\ell)$ is evaluated analytically; its expression is given in Sec.~\ref{app:pointlike_analytic}.

The IVR formulation leaves only exponentially suppressed finite-volume effects. \textcolor{revision}{We correct the dominant finite-volume effects using the difference between the ground-state model predictions in infinite volume and at the lattice volume.} More details are given in Ref.~\cite{Boyle:2025uuh} and Sec.~\ref{app:finite_volume_tensor}.
After applying this correction, the residual finite-volume effects are dominated by the $H_{ii}(t,\vec x)$ ($i=1,2,3$) component and its heavier $J^P=1^-$ intermediate states; see Sec.~\ref{app:finite_volume_tensor} for the numerical evidence from a channel-by-channel decomposition. 

The components $H_{ii}(t,\vec x)$ contribute only to the axial-vector part of the transverse-photon contribution, denoted by $\delta_{\text{Coul}}^{\text{vir,tr,A}}(m_P,m_\ell)$. 
To reduce the statistical and finite-volume errors associated with $H_{ii}(t,\vec x)$, we adopt an error-cancellation technique previously used in the real-photon-emission calculation~\cite{Boyle:2025uuh}, based on the correlation between the decay constant extracted from $H_{ii}(t,\vec x)$ and the axial-vector transverse-photon contribution. We define the ratio
\begin{equation}\label{eq:RfP}
    R_{f_P}=\frac{1}{3}\sum_{i=1}^{3}\frac{f_P^{\text{3pt},ii}}{f_P^{\text{3pt},44}}=\frac{m_P}{3}\sum_{i=1}^3\sum_{t,\vec x}H_{ii}(t,\vec x),
\end{equation}
which compares decay constants extracted from the $\mu\rho=ii$ and $\mu\rho=44$ components and approaches unity in the continuum and infinite-volume limits. We therefore construct a subtraction method equivalent to Eq.~(\ref{eq:coul_SD}):
\begin{eqnarray}
    \delta_{\text{SD},\text{Coul}}^{\text{vir}}(m_P,m_\ell)&=&\delta_{\text{Coul}}^{\text{vir,pot}}(m_P,m_\ell)+\frac{\delta_{\text{Coul}}^{\text{vir,tr,A}}(m_P,m_\ell)}{R_{f_P}}\nonumber\\
    &&\hspace{-0.7in}+\,\delta_{\text{Coul}}^{\text{vir,tr,V}}(m_P,m_\ell)-\delta_{\text{pt},\text{Coul}}^{\text{vir}}(m_P,m_\ell),\label{eq:coul_SD_R}
\end{eqnarray}
where the error cancellation is achieved through the correlation between $\delta_{\text{Coul}}^{\text{vir,tr,A}}(m_P,m_\ell)$ and $R_{f_P}$. 

\par\medskip
\noindent\textbf{Numerical analysis -}
We use the $N_f=2+1$ domain-wall fermion ensembles at the physical pion mass generated by the RBC and UKQCD collaborations~\cite{RBC:2014ntl}, comprising the 24D, 32D, 48I, and 64I ensembles.
The 24D and 32D
ensembles have the same lattice spacing but different volumes,
$m_\pi L\sim 3.3$ and $4.5$, and are used to
assess residual finite-volume effects. The 48I and 64I ensembles have
$a^{-1}=1.730$ and $2.359\,\mathrm{GeV}$ and similar volumes, and are used for the continuum
extrapolation. \textcolor{revision}{We exclude 24D and 32D from this extrapolation because they use the Iwasaki action with a dislocation-suppressing determinant ratio (DSDR), whereas 48I and 64I use the Iwasaki action.}
\textcolor{revision}{The lattice spacings are taken from Refs.~\cite{RBC:2014ntl,Blum:2023pipi,Blum:2025HLbL}, with the physical scale set by the $\Omega$-baryon mass.}
Their parameters are summarized in Table~\ref{table:ens} of Sec.~\ref{app:ensemble_results}.
\textcolor{revision}{In extracting the hadronic functions from lattice correlators, we correct for around-the-world (ATW) effects. The dominant contributions correspond to the backward propagation of the signal in the opposite time ordering and can therefore be reconstructed from lattice data. Details of the ATW corrections and their numerical tests are given in Sec.~\ref{app:atw_subtraction}.}

\begin{figure*}
    \centering
    \includegraphics[width=0.88\textwidth]{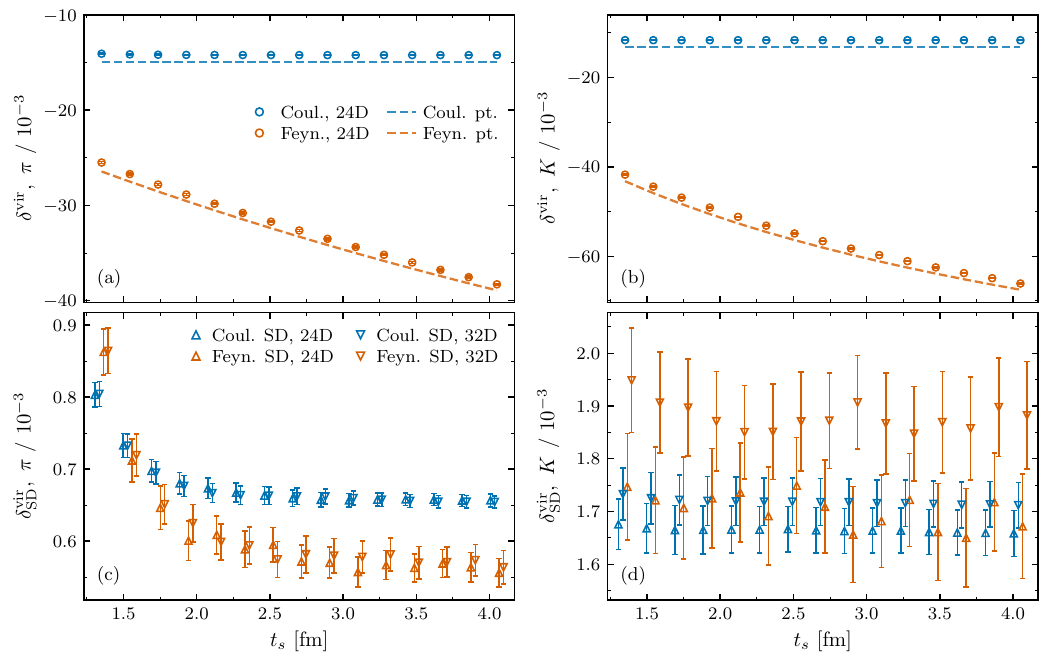}
\caption{\label{fig:M1}
Virtual correction as a function of the reconstruction time $t_s$ in the IVR method, using the 24D and 32D ensembles as examples. Upper panels: the full results before point-like subtraction for $\pi$ and $K$, showing the 24D lattice results (open circles) and the point-like contributions (dashed lines). Lower panels: the SD correction $\delta_{\text{SD}}^{\text{vir}}$ after point-like subtraction, shown for the 24D (upward triangles) and 32D (downward triangles) ensembles. Blue and orange denote the Coulomb and Feynman gauges, respectively.
}
\end{figure*}

Using 24D and 32D as representative examples, Fig.~\ref{fig:M1} compares the virtual correction as a function of $t_s$ in the two gauges, before and after subtracting the point-like contribution. All results include the finite-volume correction and the use of the $R_{f_P}$-cancellation technique. Fig.~\ref{fig:M1} shows that in both gauges, the result is dominated by the point-like contribution, but this contribution is $2$--$6$ times smaller in Coulomb gauge than in Feynman gauge. Consequently, after subtraction, the Coulomb-gauge statistical error is reduced by approximately a factor of two.

Comparing the 24D and 32D results, we find agreement within statistical errors for the pion, while \textcolor{revision}{the kaon results differ by approximately $0.9\sigma$ in Coulomb gauge and $1.3\sigma$ in Feynman gauge}. We use these differences to estimate the residual finite-volume effects.  This different behavior for the pion and kaon arises from the smaller pion mass, which makes the $H_{ii}(t,\vec{x})$ contribution to the loop integral more strongly correlated with its zero-momentum projection encoded in $R_{f_P}$, thereby making the error cancellation more effective; see Sec.~\ref{app:finite_volume_tensor} for more detailed numerical evidence.
Notably, the 24D--32D difference for the kaon is only at or below the $O(1\%)$ level in the original lattice data, but is relatively amplified after the point-like subtraction. \textcolor{revision}{We extract $\delta_{\rm SD}^{\rm vir}$ on each ensemble from constant fits over $t_s\in[3.0,3.6]\,\mathrm{fm}$ for the pion and $[2.0,3.0]\,\mathrm{fm}$ for the kaon. Further details on the selection of time-integration limits are given in Sec.~\ref{app:timecuts}.}

\begin{figure}
    \includegraphics[width=\columnwidth]{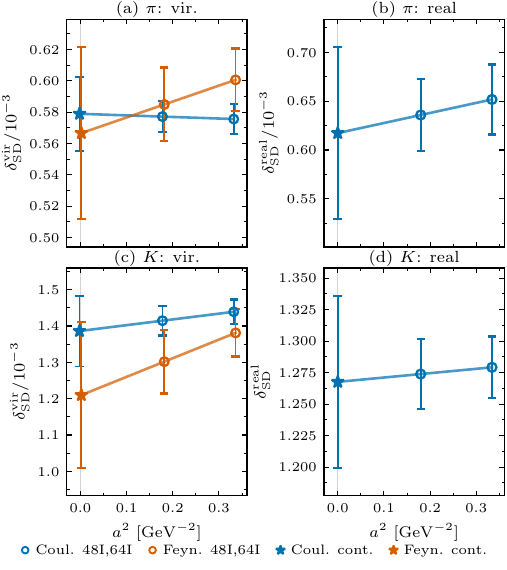}
    \caption{\label{fig:continuum} Panels (a) and (c): continuum extrapolation of the SD corrections $\delta^{\mathrm{vir}}_{\mathrm{SD}}$ for the pion and kaon, respectively, computed in both Coulomb (blue) and Feynman (orange) gauges. Panels (b) and (d): continuum extrapolation of $\delta_{\text{SD}}^{\text{real}}$ for the pion and kaon. Open circles denote the 48I and 64I results, with the 64I result corrected for the kaon-mass mismatch. Filled stars denote the continuum-extrapolated results.}
\end{figure}

Fig.~\ref{fig:continuum} shows the $a^2$-linear continuum extrapolations of $\delta_{\text{SD}}^{\text{vir}}$ and $\delta_{\text{SD}}^{\text{real}}$ using the 48I and 64I results. 
Because the unitary kaon mass on 64I is slightly heavier than that on 48I, we additionally use partially quenched calculations on 64I to correct for this kaon-mass mismatch. \textcolor{revision}{The kaon extrapolation in Fig.~\ref{fig:continuum} is performed at $m_K^{\rm 48I}=499.21\,\mathrm{MeV}$, and the final kaon results are then corrected to the physical charged-kaon mass as described in Sec.~\ref{app:meson_mass_correction}.}

\par\medskip
\noindent\textbf{Results and conclusion -}
We take the Coulomb-gauge results with the $R_{f_P}$-cancellation technique as our
final values \textcolor{revision}{because the Feynman-gauge results are more sensitive to finite-volume and ATW effects (see Secs.~\ref{app:finite_volume_tensor} and \ref{app:atw_subtraction})}. We use the Feynman-gauge results for comparison. These results, together with the ChPT predictions, are summarized in Table~\ref{tab:SD_final}. The
complete results are tabulated in Sec.~\ref{app:ensemble_results}.

The errors are estimated as follows: ``stat'' denotes the statistical uncertainty in the continuum-extrapolated result. ``FV'' denotes the residual finite-volume uncertainty, estimated from the difference between the 24D and 32D results. \textcolor{revision}{``$a^4$'' denotes the residual higher-order discretization uncertainty after the $a^2$ extrapolation. We estimate it by dimensional power counting for relative $(a\Lambda_{\rm QCD})^4$ effects with $\Lambda_{\rm QCD}=500\,\mathrm{MeV}$, giving a $0.375\%$ relative uncertainty in the continuum SD corrections. ``qdisc'' denotes the uncertainty from omitted quark-disconnected contributions. A lattice calculation of the $\pi^0\to\gamma\gamma$ transition form factor $\mathcal F_{\pi^0\gamma\gamma}(0,0)$, related by isospin symmetry to the pion vector form factor $F_V$ at the corresponding kinematic point in this work, evaluated disconnected contributions on the 24D and 64I ensembles, finding approximately $1.4\%$ and $1.3\%$ of the connected result, respectively~\cite{Lin:2024khg}. Guided by these values, we assign a $1.4\%$ relative uncertainty to the pion and kaon SD corrections.}

\textcolor{revision}{To test the compatibility of the two gauge results, we compute the difference between their continuum-extrapolated values using common bootstrap resamples to preserve statistical correlations. Defining $\Delta_{\mathrm{gauge}}^{P}=\delta_{\text{SD},\text{Coul}}^{\text{vir}}-\delta_{\text{SD},\text{Feyn}}^{\text{vir}}$ for $P=\pi,K$, we obtain $\Delta_{\mathrm{gauge}}^{\pi}=0.012(43)_{\mathrm{stat}}(14)_{\mathrm{FV}}\times10^{-3}$ and $\Delta_{\mathrm{gauge}}^{K}=0.175(149)_{\mathrm{stat}}(109)_{\mathrm{FV}}\times10^{-3}$. The FV uncertainty is estimated from the absolute 24D--32D difference in $\Delta_{\mathrm{gauge}}^{P}$. Combining the statistical and FV uncertainties in quadrature, the pion and kaon differences are $0.3\sigma$ and $0.95\sigma$ from zero, respectively, indicating that the two gauges give compatible results at the current precision.}

\begin{table}
\centering
\setlength{\tabcolsep}{3pt}
\renewcommand{\arraystretch}{1.12}
\footnotesize
\begin{tabular}{@{}lll@{}}
\hline\hline
Meson & Method & \multicolumn{1}{c}{$\delta_{\mathrm{SD}}^{\mathrm{vir}}$} \\
\hline
$\pi$ & Coul. & \textcolor{revision}{$0.579(24)_{\mathrm{stat}}(0)_{\mathrm{FV}}(2)_{a^4}(8)_{\mathrm{qdisc}}\times10^{-3}$} \\
 & Feyn. & \textcolor{revision}{$0.567(55)_{\mathrm{stat}}(14)_{\mathrm{FV}}(2)_{a^4}(8)_{\mathrm{qdisc}}\times10^{-3}$} \\
 & ChPT & $0.530(110)\times10^{-3}$ \\[2pt]
$K$ & Coul. & \textcolor{revision}{$1.359(101)_{\mathrm{stat}}(53)_{\mathrm{FV}}(5)_{a^4}(19)_{\mathrm{qdisc}}\times10^{-3}$} \\
 & Feyn. & \textcolor{revision}{$1.185(214)_{\mathrm{stat}}(162)_{\mathrm{FV}}(4)_{a^4}(17)_{\mathrm{qdisc}}\times10^{-3}$} \\
 & ChPT & $1.350(110)\times10^{-3}$ \\
\hline\hline
Meson & Method & \multicolumn{1}{c}{$\delta_{\mathrm{SD}}^{\mathrm{real}}$} \\
\hline
$\pi$ & Lattice & \textcolor{revision}{$0.617(88)_{\mathrm{stat}}(3)_{\mathrm{FV}}(2)_{a^4}(9)_{\mathrm{qdisc}}\times10^{-3}$} \\
 & ChPT & $0.730\times10^{-3}$ \\[2pt]
$K$ & Lattice & \textcolor{revision}{$1.215(70)_{\mathrm{stat}}(84)_{\mathrm{FV}}(5)_{a^4}(17)_{\mathrm{qdisc}}$} \\
 & ChPT & --- \\
\hline\hline
\end{tabular}
\caption{\label{tab:SD_final}
SD corrections $\delta_{\text{SD}}^\textrm{vir}$ and $\delta_{\text{SD}}^\textrm{real}$ for the pion and kaon. Results for $\delta_{\text{SD}}^{\text{vir}}$ are presented in both Coulomb (Coul.) and Feynman (Feyn.) gauges. 
\textcolor{revision}{The errors are statistical (stat), residual finite-volume (FV), higher-order discretization ($a^4$), and quark-disconnected (qdisc) uncertainties.} The results are compared with those from ChPT~\cite{Cirigliano:2007xi}. \textcolor{revision}{The kaon real-photon correction is not included in $R_{e/\mu}$.}
}
\end{table}

Combining the SD correction $\Delta_{\alpha,\text{SD}}$ computed in Coulomb gauge with the known point-like contribution $\Delta_{\alpha,\text{pt}}$ and the leading-log result for the higher-order QED contribution $\Delta_{\alpha^{n\geq2}}$~\cite{Marciano:1993sh}, we obtain for the pion,
\begin{equation}\label{eq:R_results}
R_{e/\mu} = \textcolor{revision}{1.23516(9)_{\mathrm{stat}}(4)_{\alpha^{n\geq 2}}(1)_{a^4}(2)_{\mathrm{qdisc}}}\times 10^{-4},
\end{equation}
and for the kaon,
\begin{equation}\label{eq:R_results2}
R_{e/\mu} = \textcolor{revision}{2.47661(26)_{\mathrm{stat}}(14)_{\mathrm{FV}}(8)_{\alpha^{n\geq 2}}(1)_{a^4}(5)_{\mathrm{qdisc}}}\times10^{-5}.
\end{equation}
For the pion, the finite-volume uncertainty estimated from the 24D–32D difference is smaller than the last quoted digit and is therefore omitted.
The higher-order QED error, denoted by ``$\alpha^{n\geq2}$'', is estimated as $0.003\%$ by combining
an estimate of the size of the next-to-leading-logarithmic (NLL) terms, inferred from the difference between one- and two-loop QED
running, with an estimate of the non-log-enhanced
$\mathcal{O}(\alpha^2)$ terms. The detailed estimate is given in Sec.~\ref{app:higher_order_qed}. 

\begin{figure}
    \centering
    \includegraphics[width=0.9\columnwidth]{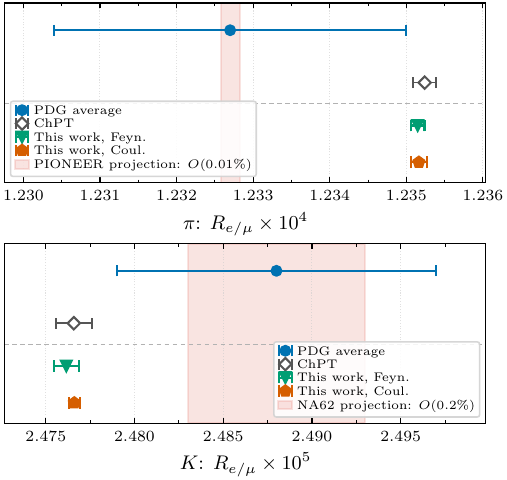}
    \caption{\label{fig:Rratio} $R_{e/\mu}$ results for the pion (top) and kaon (bottom) from this work, compared with the current experimental averages~\cite{ParticleDataGroup:2024cfk} and the ChPT predictions~\cite{Cirigliano:2007xi,Cirigliano:2007ga}. The experimental results are dominated by the PIENU measurement for the pion~\cite{PiENu:2015seu} and the NA62 measurement for the kaon~\cite{NA62:2012lny}. The vertical shaded bands indicate the projected precisions of the PIONEER experiment~\cite{PIONEER:2022yag,PIONEER:2022alm,PIONEER:2025idw} and the updated NA62 measurement~\cite{Massri:2025NA62}. \textcolor{revision}{For our pion result, the total uncertainty is slightly smaller in Feynman gauge. This is because the statistical error is dominated by real photons, and adding the virtual- and real-photon contributions with correlations retained reduces the combined statistical error, with a larger reduction in Feynman gauge.}}
\end{figure}

In Fig.\,\ref{fig:Rratio}, we compare our results with those from ChPT, the current experimental averages, and the projected precisions of future experimental measurements. \textcolor{revision}{The lattice results agree with the ChPT predictions for both mesons. The kaon ChPT uncertainty is significantly larger because} Ref.~\cite{Cirigliano:2007xi} conservatively inflated the higher-order ChPT error by a factor of $4$. The uncertainty in the lattice calculation is comparable to the future PIONEER projected precision~\cite{PIONEER:2022yag,PIONEER:2022alm,PIONEER:2025idw}. The NA62 projected precision of about $0.2\%$ for the kaon ratio~\cite{Massri:2025NA62} remains above the theoretical uncertainty.

As the first lattice determination of the SD correction to $R_{e/\mu}$, this work reduces the associated hadronic uncertainty and yields the most precise Standard-Model prediction to date. Our results establish a first-principles Standard-Model benchmark that will be crucial for interpreting future high-precision tests of LFU, especially when experiments such as PIONEER reach comparable precision.

\par\medskip
\noindent\textbf{Acknowledgments -}
We would like to thank our colleagues in the RBC and UKQCD Collaborations for helpful discussions and support.
P.B. and T.I. were supported in part by US DOE Contract DESC0012704(BNL) and the Scientific Discovery
through Advanced Computing (SciDAC) program LAB
22-2580. X.F. has been supported in part by NSFC of China under Grant No. 12125501 and Grant No. 12550007. 
L.J. acknowledges the support of DOE Office of Science Early Career Award DE-SC0021147 and DOE
grants DE-SC0010339 and DE-SC0026314. C.T.S. is partially supported by STFC consolidated grant ST/X000583/1.
X.Y.T. was supported by US DOE Contract
DESC0012704(BNL).
The research reported in this work made use of computing and long-term storage
facilities of the USQCD Collaboration, which are funded by the Office of Science
of the U.S. Department of Energy.
An award of computer time was provided by the ASCR Leadership Computing Challenge (ALCC) program. This research used resources of the Argonne Leadership Computing Facility, which is a U.S. Department of Energy Office of Science User Facility operated under contract DE-AC02-06CH11357.

\bibliography{reference}


\clearpage
\onecolumngrid
\begin{center}
  \textbf{\large Supplemental Material}\\[5pt]
  {\small First Lattice Calculation of Structure-Dependent Radiative Corrections to Lepton-Flavor-Universality Ratios in Leptonic Decays}\\[5pt]
  {\small Peter Boyle, Norman H. Christ, Xu Feng, Taku Izubuchi, Luchang Jin, Christopher T. Sachrajda, and Xin-Yu Tuo}
\end{center}
\twocolumngrid
\setcounter{secnumdepth}{3}
\setcounter{section}{0}
\setcounter{equation}{0}
\setcounter{figure}{0}
\setcounter{table}{0}
\renewcommand{\thesection}{S\arabic{section}}
\renewcommand{\thesubsection}{\thesection.\arabic{subsection}}
\renewcommand{\theequation}{S\arabic{equation}}
\renewcommand{\thefigure}{S\arabic{figure}}
\renewcommand{\thetable}{S\arabic{table}}
\makeatletter\renewcommand{\p@subsection}{}\makeatother

\section{Derivation of IVR formulas}
\label{app:virtual_ivr}
This section derives the scalar-function representation used to evaluate the tensor convolutions in Eq.~\eqref{eq:delta_coul_lat}. Sec.~\ref{app:coulomb_derivation} gives the Coulomb-gauge derivation, while Sec.~\ref{app:feynman_crosscheck} presents the corresponding Feynman-gauge expressions.

\subsection{Coulomb-gauge IVR formula}
\label{app:coulomb_derivation}
In Coulomb gauge, the virtual correction in Eq.~(\ref{eq:delta_vir_def}) separates into Coulomb-potential and transverse-photon contributions,
\begin{equation}\label{delta_TB_M_Coulomb}
\begin{aligned}
    &\delta_{\text{Coul}}^{\text{vir,pot}}(m_P,m_\ell)=i\frac{2\pi\alpha}{f_P m_P^2 m_\ell^2(1-r_\ell)}\\
    &\quad\times\int\frac{d^4k}{(2\pi)^4}\frac{L_{\mu 0}(k,p,p_\ell)\, H_M^{\mu 0}}{\vec{k}^2\left((p_\ell-k)^2-m_\ell^2+i\epsilon\right)},\\
    &\delta_{\text{Coul}}^{\text{vir,tr}}(m_P,m_\ell)=i\frac{2\pi\alpha}{f_P m_P^2 m_\ell^2(1-r_\ell)}\\
    & \times \int\frac{d^4k}{(2\pi)^4}\frac{\sum_{\lambda=1,2} L_{\mu}^{\sigma}(k,p,p_\ell)\, \epsilon_\sigma(k,\lambda)\epsilon^*_\rho(k,\lambda)\, H_M^{\mu\rho}(k,p)}{\left(k^2+i\epsilon\right)\left((p_\ell-k)^2-m_\ell^2+i\epsilon\right)}.
\end{aligned}
\end{equation}

We use the scalar-function decomposition introduced in Refs.~\cite{Tuo:2021ewr,Boyle:2025uuh} to improve computational efficiency and reduce data-storage requirements. We first project the Minkowski hadronic matrix element $H_M^{\mu\rho}(k,p)$ onto six Lorentz-tensor structures,
\begin{equation}\label{eq:scalarproj}
\begin{aligned}
    \tilde I_i(\rho_1,\rho_2)&=\tilde l_{i,\mu\rho}(k,p)\,H_M^{\mu\rho}(k,p),\\
    \tilde l_{i,\mu\rho}&=\big\{g_{\mu\rho}p^2,\ p_\mu p_\rho,\ p_\mu k_\rho,\\
    &\qquad k_\mu p_\rho,\ k_\mu k_\rho,\ i\varepsilon_{\mu\rho\alpha\beta}k^\alpha p^\beta\big\},
\end{aligned}
\end{equation}
which depend on $\rho_1=k^2/m_P^2$ and $\rho_2=(p-k)^2/m_P^2$. In terms of these scalar functions, the hadronic matrix element can be decomposed as
\begin{equation}\label{eq:scalardecomp}
    H_M^{\mu\rho}(k,p)=\sum_{i=1}^6\omega_i^{\mu\rho}(k,p)\,\tilde I_i(\rho_1,\rho_2),
\end{equation}
with the known coefficients $\omega_i^{\mu\rho}(k,p)$. 

The Coulomb-potential and transverse-photon contributions can be written in the scalar-function form as
\begin{equation}\label{eq:delta_coul_scalar}
\begin{aligned}
    &\delta_{\text{Coul}}^{\text{vir,pot}}(m_P,m_\ell)=i\frac{2\pi\alpha}{f_P m_P^2}\\
    &\quad\times\int\frac{d^4k}{(2\pi)^4}\frac{\sum_{i=1}^6 \bar L^{\text{pot}}_i(\rho_1,\rho_2,\theta)\,\tilde I_i(\rho_1,\rho_2)}{\vec k^2\big((p_\ell-k)^2-m_\ell^2+i\epsilon\big)},\\
    &\delta_{\text{Coul}}^{\text{vir,tr}}(m_P,m_\ell)=i\frac{2\pi\alpha}{f_P m_P^2}\\
    &\quad\times\int\frac{d^4k}{(2\pi)^4}\frac{\sum_{i=1}^6\bar L^{\text{tr}}_i(\rho_1,\rho_2,\theta)\,\tilde I_i(\rho_1,\rho_2)}{\big(k^2+i\epsilon\big)\big((p_\ell-k)^2-m_\ell^2+i\epsilon\big)},
\end{aligned}
\end{equation}
where the dimensionless leptonic scalar structures are defined by
\begin{equation}\label{eq:Lbar_coul}
\begin{aligned}
    \bar L^{\text{pot}}_i(\rho_1,\rho_2,\theta)&=\frac{L_{\mu0}(k,p,p_\ell)\,\omega_i^{\mu0}(k,p)}{m_\ell^2(1-r_\ell)},\\
    \bar L^{\text{tr}}_i(\rho_1,\rho_2,\theta)&=\frac{1}{m_\ell^2(1-r_\ell)}\sum_{\lambda=1,2}\varepsilon_\sigma(k,\lambda)\varepsilon^*_\rho(k,\lambda)\\
    &\quad\times L_\mu^{\sigma}(k,p,p_\ell)\,\omega_i^{\mu\rho}(k,p),
\end{aligned}
\end{equation}
and depend on $\rho_1$, $\rho_2$, and the angle $\theta$ between the photon momentum $\vec k$ and the final-state lepton momentum $\vec p_\ell$.

For the coordinate-space implementation, we introduce the six scalar
functions constructed from the hadronic function $H_{\mu\nu}(t,\vec{x})$ defined in Eq.~\eqref{IE}:
\begin{align}
I_1(t,|\vec{x}|)&=\delta_{\mu\nu}H_{\mu\nu}(t,\vec{x}),
\nonumber\\
I_2(t,|\vec{x}|)&=H_{44}(t,\vec{x}),
\nonumber\\
I_3(t,|\vec{x}|)&=x_mH_{4m}(t,\vec{x}),
\nonumber\\
I_4(t,|\vec{x}|)&=x_mH_{m4}(t,\vec{x}),
\nonumber\\
I_5(t,|\vec{x}|)&=x_mx_nH_{mn}(t,\vec{x}),
\nonumber\\
I_6(t,|\vec{x}|)&=-\epsilon_{mnk}x_kH_{mn}(t,\vec{x}).
\label{eq:scalar_lattice_inputs}
\end{align}
Here $m,n\in\{1,2,3\}$ denote spatial directions.
The functions $\tilde I_i(\rho_1,\rho_2)$ are related to $I_j(t,|\vec x|)$ by the IVR formula
\begin{equation}\label{eq:scalar_calc}
\begin{aligned}
    &\tilde I_i(\rho_1,\rho_2)\\
    =&\,f_Pm_P^3\sum_{j=1}^6\int d^3x\int_{-t_s}^{\infty}dt \,e^{k^0t}\phi_{ij}(\rho_1,\rho_2;|\vec x|)\,I_j(t,|\vec x|)\\
    +&\,f_Pm_P^3\sum_{j=1}^6\int d^3x\,\frac{e^{-k^0t_s}\phi_{ij}(\rho_1,\rho_2;|\vec x|)\,I_j(-t_s,|\vec x|)}{k^0+E_P(\vec k)-m_P},
\end{aligned}
\end{equation}
where the on-shell intermediate-state energy is $E_P(\vec k)=\sqrt{\vec k^2+m_P^2}$. The first term gives the short-distance contribution from $t>-t_s$, while the second gives the long-distance contribution from $t\le-t_s$, reconstructed from the hadronic function at $t=-t_s$.
The weight matrix $\phi_{ij}(\rho_1,\rho_2;|\vec x|)$, obtained by averaging the Fourier factor $e^{-i\vec k\cdot\vec x}$ over the photon-momentum direction (see Ref.~\cite{Boyle:2025uuh} for the derivation), is given by
\begin{widetext}
\begin{equation}\label{eq:phi}
    \phi_{ij}(\rho_1,\rho_2;|\vec x|)=
    \begin{pmatrix}
        j_0 & 0 & 0 & 0 & 0 & 0\\
        0 & j_0 & 0 & 0 & 0 & 0\\
        0 & \tfrac{y}{2}j_0 & -\tfrac{m_Pz^2}{4}\tfrac{j_1}{\varphi} & 0 & 0 & 0\\
        0 & \tfrac{y}{2}j_0 & 0 & -\tfrac{m_Pz^2}{4}\tfrac{j_1}{\varphi} & 0 & 0\\
        -\tfrac{z^2}{4}\tfrac{j_1}{\varphi} & \tfrac{y^2}{4}j_0+\tfrac{z^2}{4}\tfrac{j_1}{\varphi} & -\tfrac{m_Pyz^2}{8}\tfrac{j_1}{\varphi} & -\tfrac{m_Pyz^2}{8}\tfrac{j_1}{\varphi} & \tfrac{m_P^2z^4}{16}\tfrac{j_2}{\varphi^2} & 0\\
        0 & 0 & 0 & 0 & 0 & \tfrac{m_Pz^2}{4}\tfrac{j_1}{\varphi}
    \end{pmatrix},
\end{equation}
\end{widetext}
where $y=2k^0/m_P=1+\rho_1-\rho_2$ and $z=2|\vec k|/m_P=\sqrt{1+\rho_1^2+\rho_2^2-2\rho_1-2\rho_2-2\rho_1\rho_2}$. Here, $j_0$, $j_1$, and $j_2$ denote the spherical Bessel functions evaluated at $\varphi=|\vec k||\vec x|=z m_P|\vec x|/2$.

\begin{figure}
    \centering
    \includegraphics[width=0.48\columnwidth]{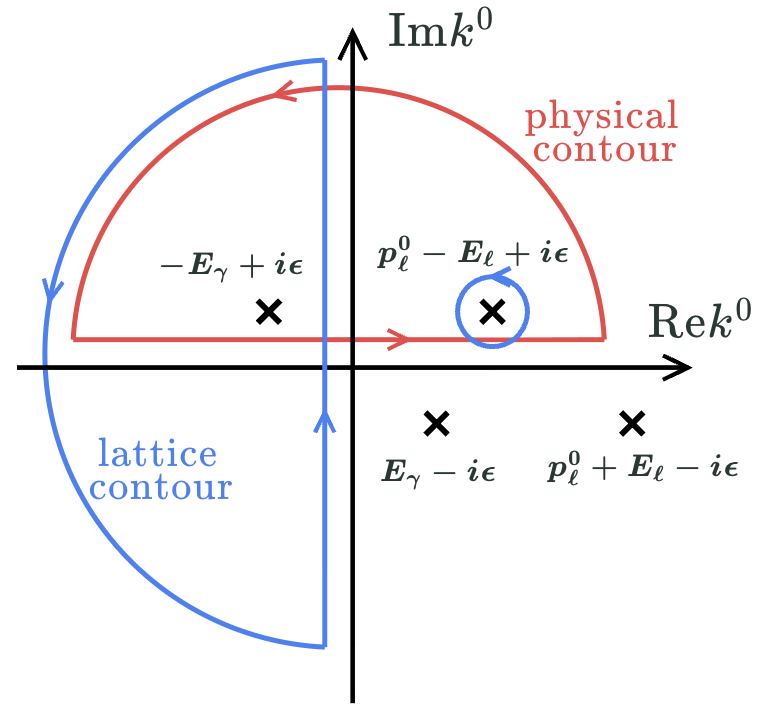}\hfill
    \includegraphics[width=0.48\columnwidth]{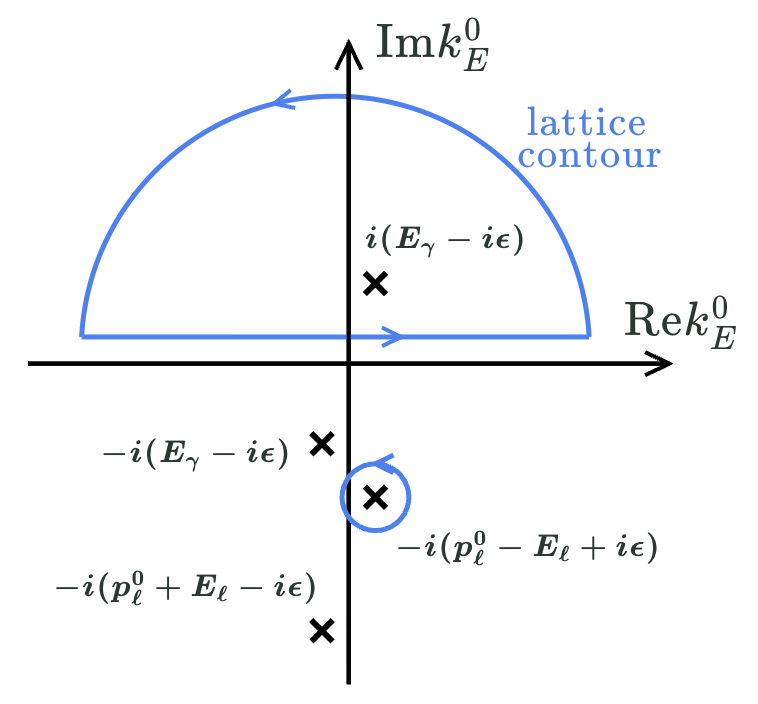}
    \caption{\label{fig:Wick}Wick rotation used in the lattice calculation of the virtual-photon loop. Left: the complex $k^0$ plane, showing the singularities of the photon and lepton propagators, the physical Minkowski contour (red), and the deformed lattice contour $C_{\text{lat}}$ (blue). Right: the same lattice contour $C_{\text{lat}}$ in the complex $k_E^0$ plane.}
\end{figure}

We Wick rotate ($k^0\to ik_E^0$) the loop integral in Eq.~(\ref{eq:delta_coul_scalar}) to Euclidean space. Fig.~\ref{fig:Wick} illustrates the contour deformation, using the $t>0$ time ordering as an example. In the left panel, the singularities of the lepton and photon propagators in the complex $k^0$ plane are shown together with the Minkowski contour (red). The Euclidean contour $C_{\text{lat}}$ used in the lattice calculation (blue) must be deformed so as to enclose the residues of the appropriate physical singularities. The right panel shows the same contour $C_{\text{lat}}$ in the complex $k_E^0$ plane. 

The tensor convolutions in Eq.~\eqref{eq:delta_coul_lat} are evaluated in scalar form as
\begin{equation}
\begin{aligned}
\delta_{\text {Coul}}^{\mathrm{vir}, \text {pot}}(m_P,m_\ell) & =\int d^3 x \int_{-t_s}^{\infty} d t \sum_j I_j(t,|\vec{x}|) \bar{\mathcal{K}}_j^{\mathrm{pot}}(t,|\vec{x}|) \\
+&\int d^3 x \sum_j I_j\left(-t_s,|\vec{x}|\right) \bar{\mathcal{K}}_j^{(l), \mathrm{pot}}\left(-t_s,|\vec{x}|\right), \\
\delta_{\text {Coul}}^{\mathrm{vir}, \mathrm{tr}}(m_P,m_\ell) & =\int d^3 x \int_{-t_s}^{\infty} d t \sum_j I_j(t,|\vec{x}|) \bar{\mathcal{K}}_j^{\mathrm{tr}}(t,|\vec{x}|).
\end{aligned}
\end{equation}
The transverse-photon contribution does not admit a $P$-meson intermediate state and therefore requires no temporal reconstruction.

Applying the Wick rotation with the deformed lattice contour to Eq.~(\ref{eq:delta_coul_scalar}) and combining it with Eq.~(\ref{eq:scalar_calc}) yields the following weight functions
\begin{widetext}
\begin{equation}\label{weightfunction_TB_short_long_coul}
    \begin{aligned}
        \bar{\mathcal{K}}^{\text{pot}}_{j}(t,|\vec{x}|)&=2\pi\alpha m_P \int\frac{d^3k}{(2\pi)^3}\int_{C_{\text{lat}}}\frac{dk_E^0}{2\pi} e^{ik_E^0 t}\frac{\sum_{i=1}^6 \bar{L}^{\text{pot}}_i(\rho_1,\rho_2,\theta) \phi_{ij}(\rho_1,\rho_2;|\vec{x}|)}{E_\gamma(\vec{k})^2((p_{\ell,E}^0-k_E^0)^2+E_\ell(\vec{k})^2-i\epsilon)},\\
        \bar{\mathcal{K}}^{(l),\text{pot}}_{j}(-t_s,|\vec{x}|)&=2\pi\alpha m_P \int\frac{d^3k}{(2\pi)^3}\int_{C_{\text{lat}}}\frac{dk_E^0}{2\pi}\frac{e^{-ik_E^0 t_s}}{ik_E^0+E_P(\vec{k})-m_P}\frac{\sum_{i=1}^6 \bar{L}^{\text{pot}}_i(\rho_1,\rho_2,\theta) \phi_{ij}(\rho_1,\rho_2;|\vec{x}|)}{E_\gamma(\vec{k})^2((p_{\ell,E}^0-k_E^0)^2+E_\ell(\vec{k})^2-i\epsilon)},\\
        \bar{\mathcal{K}}^{\text{tr}}_{j}(t,|\vec{x}|)&=-2\pi\alpha m_P \int\frac{d^3k}{(2\pi)^3}\int_{C_{\text{lat}}}\frac{dk_E^0}{2\pi} e^{ik_E^0 t}\frac{\sum_{i=1}^6 \bar{L}^{\text{tr}}_i(\rho_1,\rho_2,\theta) \phi_{ij}(\rho_1,\rho_2;|\vec{x}|)}{((k_E^0)^2+E_\gamma(\vec{k})^2-i\epsilon)((p_{\ell,E}^0-k_E^0)^2+E_\ell(\vec{k})^2-i\epsilon)},
    \end{aligned}
\end{equation}
\end{widetext}
where the Euclidean momenta are $k_E=(-ik^0,\vec{k})$, $p_{\ell,E}=(-ip_{\ell}^0,\vec{p}_\ell)$, and $p_{E}=(-im_P,\vec{0})$, and the photon and lepton energies are $E_\gamma(\vec k)=|\vec k|$ and $E_\ell(\vec k)=\sqrt{\vec k^2+m_\ell^2}$. The quantities $\bar{L}^{\text{pot}}_i$ and $\bar{L}^{\text{tr}}_i$ are defined in Eq.~(\ref{eq:Lbar_coul}), with their arguments expressed in terms of Euclidean momenta as $\rho_1=-(k_E)^2/m_P^2$ and $\rho_2=-(p_{E}-k_E)^2/m_P^2$. The numerical evaluation of these weight functions encounters difficulties from the collinear singularity for a final-state electron and from the $|\vec{k}|\to\infty$ region at $t=0$; their treatment is described in Sec.~\ref{app:weight_functions}.

\subsection{Feynman-gauge IVR formula}
\label{app:feynman_crosscheck}
For completeness, we summarize the corresponding Feynman-gauge IVR formula. The derivation in Feynman gauge closely parallels that in Coulomb gauge. 

In Feynman gauge, the virtual correction can be written in scalar-function form as
\begin{equation}\label{eq:delta_feyn_scalar}
\begin{aligned}
    &\delta^{\text{vir}}_{\text{Feyn}}(m_P,m_\ell)=-i\frac{2\pi\alpha}{f_P m_P^2}\\
    &\quad\times\int\frac{d^4k}{(2\pi)^4}\frac{\sum_{i=1}^6 \bar L_i^{\text{Feyn}}(\rho_1,\rho_2,\theta)\,\tilde I_i(\rho_1,\rho_2)}{(k^2+i\epsilon)\big((p_\ell-k)^2-m_\ell^2+i\epsilon\big)},
\end{aligned}
\end{equation}
where the dimensionless leptonic scalar functions are defined by
\begin{equation}
    \bar L^{\text{Feyn}}_i(\rho_1,\rho_2,\theta)=\frac{L_{\mu\rho}(k,p,p_\ell)\omega_i^{\mu\rho}(k,p)}{m_\ell^2(1-r_\ell)}.
\end{equation}

Unlike in Coulomb gauge, the Feynman-gauge computation contains an IR divergence. To remove this divergence in the lattice calculation, we perform an IR subtraction within the IVR scheme and define the IR-convergent part of the scalar functions as
\begin{equation}\label{eq:scalar_calc_sub}
\begin{aligned}
    &\tilde I^{\text{con}}_i(\rho_1,\rho_2;t_s)\\
    =&f_P\,m_P^3\sum_{j}\int d^3x\int_{-t_s}^{\infty}dt\,e^{k^0 t} \phi_{ij}(\rho_1,\rho_2;|\vec x|)\,I_j(t,|\vec x|)\\
    +&f_P\,m_P^3\sum_{j}\int d^3x\,\frac{e^{-k^0 t_s}}{k^0+E_P(\vec k)-m_P}\\
    &\times\Big(\phi_{ij}(\rho_1,\rho_2;|\vec x|)-\phi^{\text{div}}_{ij}(\rho_1,\rho_2)\Big) I_j(-t_s,|\vec x|),
\end{aligned}
\end{equation}
where the IR subtraction in the long-distance region is implemented through
\begin{equation}\label{eq:phidiv}
    \phi^{\text{div}}_{ij}(\rho_1,\rho_2)=
    \begin{pmatrix}
        1 & 0 & 0 & 0 & 0 & 0\\
        0 & 1 & 0 & 0 & 0 & 0\\
        0 & \tfrac{y}{2} & 0 & 0 & 0 & 0\\
        0 & \tfrac{y}{2} & 0 & 0 & 0 & 0\\
        -\tfrac{z^2}{12} & \tfrac{y^2}{4}+\tfrac{z^2}{12} & 0 & 0 & 0 & 0\\
        0 & 0 & 0 & 0 & 0 & 0
    \end{pmatrix},
\end{equation}
which is obtained by taking the $\vec{k}\to 0$ limit of the $j=1,2$ columns of $\phi_{ij}$ in Eq.~(\ref{eq:phi}). This subtraction is equivalent to subtracting the IR-divergent part of the hadronic matrix element,
\begin{equation}\label{eq:HMdiv}
    H_{M,\text{div}}^{\mu\rho}(k,p;t_s)=f_P m_P\frac{e^{-k^0 t_s}}{k^0+E_P(\vec k)-m_P}\delta_{\mu 0}\delta_{\rho 0},
\end{equation}
which contains no hadronic SD information.

After performing the same Wick rotation as in Coulomb gauge, the lattice formula for the IR-subtracted virtual-photon loop in Feynman gauge reads
\begin{equation}\label{eq:delta_feyn_lat}
\begin{aligned}
    &\delta^{\text{vir,con}}_{\text{Feyn}}(m_P,m_\ell;t_s)\\
    =&\int d^3x\!\int_{-t_s}^{\infty}\!dt\sum_j I_j(t,|\vec x|)\bar{\mathcal{K}}_j(t,|\vec x|)\\
    +&\int d^3x\sum_j I_j(-t_s,|\vec x|)\bar{\mathcal{K}}^{(l),\text{con}}_j(-t_s,|\vec x|),
\end{aligned}
\end{equation}
where the short-distance weight function and the IR-subtracted long-distance weight function are given by
\begin{widetext}
\begin{equation}\label{weightfunction_TB_short_long}
    \begin{aligned}
        \bar{\mathcal{K}}_{j}(t,|\vec{x}|)&=2\pi\alpha m_P \int\frac{d^3k}{(2\pi)^3}\int_{C_{\text{lat}}}\frac{dk_E^0}{2\pi} e^{ik_E^0 t}\\
        &\times\frac{\sum_{i=1}^6 \bar{L}^{\text{Feyn}}_i(\rho_1,\rho_2,\theta) \phi_{ij}(\rho_1,\rho_2;|\vec{x}|)}{((k_E^0)^2+E_\gamma(\vec{k})^2-i\epsilon)((p_{\ell,E}^0-k_E^0)^2+E_\ell(\vec{k})^2-i\epsilon)},\\
        \bar{\mathcal{K}}^{(l),\text{con}}_{j}(-t_s,|\vec{x}|)&=2\pi\alpha m_P \int\frac{d^3k}{(2\pi)^3}\int_{C_{\text{lat}}}\frac{dk_E^0}{2\pi}\frac{e^{-ik_E^0 t_s}}{ik_E^0+E_P(\vec{k})-m_P}\\ &\times \frac{\sum_{i=1}^6 \bar{L}^{\text{Feyn}}_i(\rho_1,\rho_2,\theta) \left(\phi_{ij}(\rho_1,\rho_2;|\vec{x}|)-\phi_{ij}^{\text{div}}(\rho_1,\rho_2)\right)}{((k_E^0)^2+E_\gamma(\vec{k})^2-i\epsilon)((p_{\ell,E}^0-k_E^0)^2+E_\ell(\vec{k})^2-i\epsilon)}.
    \end{aligned}
\end{equation}
\end{widetext}

We then extract the SD contribution from Eq.~(\ref{eq:delta_feyn_lat}) as
\begin{equation}\label{calc_Feyn}
\begin{aligned}
    \delta^{\text{vir}}_{\text{SD},\text{Feyn}}(m_P,m_\ell;t_s)&=\delta^{\text{vir},\text{con}}_{\text{Feyn}}(m_P,m_\ell;t_s)\\
    &-\delta^{\text{vir},\text{con}}_{\text{pt},\text{Feyn}}(m_P,m_\ell;t_s),
\end{aligned}
\end{equation}
where the point-like contribution to the IR-convergent part, $\delta^{\text{vir},\text{con}}_{\text{pt},\text{Feyn}}(m_P,m_\ell;t_s)$, can be evaluated analytically; its expression is given in Sec.~\ref{app:pointlike_analytic}.

As in Coulomb gauge, the $R_{f_P}$-cancellation technique can be used to reduce the statistical and finite-volume errors associated with the component $H_{ii}(t,\vec{x})$ ($i=1,2,3$). With this technique, the SD contribution is given by
\begin{equation}\label{calc_Feyn_R}
\begin{aligned}
    \delta^{\text{vir}}_{\text{SD},\text{Feyn}}(m_P,m_\ell;t_s)&=\delta^{\text{vir},\text{con},\overline{ii}}_{\text{Feyn}}(m_P,m_\ell;t_s)\\
    &+\frac{1}{R_{f_P}}\delta^{\text{vir},\text{con},ii}_{\text{Feyn}}(m_P,m_\ell;t_s)\\
    &-\delta^{\text{vir},\text{con}}_{\text{pt},\text{Feyn}}(m_P,m_\ell;t_s).
\end{aligned}
\end{equation}
Here, $\delta^{\text{vir},\text{con},ii}_{\text{Feyn}}(m_P,m_\ell;t_s)$ denotes the contribution from $H_{ii}(t,\vec{x})$ ($i=1,2,3$), while $\delta^{\text{vir},\text{con},\overline{ii}}_{\text{Feyn}}(m_P,m_\ell;t_s)$ denotes the contribution from all other components.

\section{Numerical evaluation of the weight functions}
\label{app:weight_functions}
Here, we describe the numerical evaluation of the weight functions derived in Sec.~\ref{app:virtual_ivr}. For these functions, we first perform the $k_E^0$ integration analytically along the contour $C_{\text{lat}}$. Fig.~\ref{fig:Wick} shows this contour for the $t>0$ time ordering, where it lies in the upper half of the complex $k_E^0$ plane; for $t<0$, $C_{\text{lat}}$ lies in the lower half plane and is deformed analogously to enclose the appropriate physical singularities. After the analytic $k_E^0$ integration, the remaining integral over the spatial loop momentum $\vec{k}$ depends only on its modulus $|\vec{k}|$ and on the angle $\theta$ between $\vec{k}$ and the final-state lepton momentum $\vec{p}_\ell$. It is therefore a two-dimensional integral,
\(
    \int d^3 k=2\pi\int_0^\infty |\vec{k}|^2 d|\vec{k}|\int_{-1}^{1}dc,~c\equiv\cos\theta
\).
A direct numerical evaluation of this two-dimensional integral encounters two difficulties:
\begin{itemize}
    \item At $t=0$, the numerical integration is challenging in the $|\vec{k}|\to\infty$ region. This is because the integrand has no exponential suppression from $e^{-k^0|t|}$ at $t=0$, while the spherical Bessel functions in $\phi_{ij}(\rho_1,\rho_2;|\vec{x}|)$ oscillate rapidly as $\varphi=|\vec{k}||\vec{x}|$ grows.
    \item For $m_\ell=m_e$, the integral suffers from a near-collinear singularity when the photon momentum $\vec{k}$ is nearly collinear with the final-state lepton momentum $\vec{p}_\ell$. Although the physical electron mass $m_e$ is nonzero and therefore no true collinear divergence occurs in the $c\to 1$ region, the integrand there is much larger than in other regions, making direct numerical integration inefficient.
\end{itemize}
Below, we describe our treatment of these two difficulties, focusing on the main ideas rather than the explicit formulas used in practice.

\subsection{$t=0$ integral: large-momentum expansion of the integrand}
The ultraviolet (UV) divergence in the virtual-photon loop considered here cancels between the $e$ and $\mu$ channels. As a result, the contribution from the point $t=0$, $\vec{x}=\vec{0}$ vanishes as the lattice spacing decreases, corresponding to an $O(a^2)$ effect, and is removed automatically in the $a\to 0$ continuum extrapolation. We therefore set the weight function at this point to zero. For $t=0$ and $\vec{x}\neq\vec{0}$, all integrals entering the weight functions can be written in the common form
\begin{equation}
\begin{aligned}
    f(t=0,|\vec{x}|)=&\int_0^\infty |\vec{k}|^2 d|\vec{k}| \int_{-1}^{1} dc \Big(g_0(|\vec{k}|,c) j_0(\varphi)\\
    &+g_1(|\vec{k}|,c) \frac{j_1(\varphi)}{\varphi}+g_2(|\vec{k}|,c) \frac{j_2(\varphi)}{\varphi^2} \Big),
\end{aligned}
\end{equation}
where $g_i(|\vec{k}|,c)$ falls off at large $|\vec{k}|$ as $O(1/|\vec{k}|^{n})$ with $n\geq 2$. The power-law falloff is not sufficient to damp the rapid oscillations of the spherical Bessel functions in the $|\vec{k}|\to\infty$ region, leading to poor convergence of the two-dimensional numerical integration.

To address this issue, we expand the coefficients $g_i(|\vec{k}|,c)$ in powers of $1/|\vec{k}|$ as
\begin{equation}
\begin{aligned}
    g_i(|\vec{k}|,c)&=\bar{g}_{i,N}(|\vec{k}|,c)+O(1/|\vec{k}|^{N+1}),\\
    \bar{g}_{i,N}(|\vec{k}|,c)&=\sum_{n\leq N} \frac{\tilde{g}_{i,n}(c)}{|\vec{k}|^n},
\end{aligned}
\end{equation}
where $\bar{g}_{i,N}(|\vec{k}|,c)$ denotes the expansion of $g_i(|\vec{k}|,c)$ through $N$th order in $1/|\vec{k}|$ in the large-$|\vec{k}|$ limit. We split the integral into a low-momentum region, $\int_0^{k_\text{cut}} d|\vec{k}|\cdots$, where the large-$|\vec{k}|$ expansion is not applicable, and a high-momentum region, $\int_{k_\text{cut}}^\infty d|\vec{k}|\cdots$, where it can be used. The low-momentum contribution is evaluated directly by numerical integration. In the high-momentum region, replacing $g_i(|\vec{k}|,c)$ by its expansion $\bar{g}_{i,N}(|\vec{k}|,c)$ yields an analytically integrable contribution. The remaining correction from $g_i(|\vec{k}|,c)-\bar{g}_{i,N}(|\vec{k}|,c)$, which falls off as $O(1/|\vec{k}|^{N+1})$, is then evaluated numerically. We validate the procedure by varying the cutoff $k_\text{cut}$ and confirming the stability of the final result.

\subsection{Near-collinear enhancement for $m_\ell=m_e$}
The Feynman-gauge weight functions [Eq.~(\ref{weightfunction_TB_short_long})] and the Coulomb-gauge transverse-photon weight functions [Eq.~(\ref{weightfunction_TB_short_long_coul})] both exhibit a near-collinear enhancement for $m_\ell=m_e$. 
The integral structure affected by the collinear singularity can be cast into the generic form
\begin{equation}
    f(t,|\vec{x}|)=\int_0^\infty |\vec{k}|^2 d|\vec{k}| \int_{-1}^1 dc \,\frac{h(|\vec{k}|,c;t,|\vec{x}|)}{c_*-c},
\end{equation}
where $c_*=(1+r_\ell)/(1-r_\ell)$. To improve the numerical integration, we subtract the near-collinear structure as
\begin{equation}
\begin{aligned}
    f(t,|\vec{x}|)&=\int_0^\infty |\vec{k}|^2 d|\vec{k}| \int_{-1}^{c_{\text{cut}}} dc\,\frac{h(|\vec{k}|,c;t,|\vec{x}|)}{c_*-c}\\
    &+\int_0^\infty |\vec{k}|^2 d|\vec{k}| \int_{c_{\text{cut}}}^1 dc\\
    &\quad\times\frac{h(|\vec{k}|,c;t,|\vec{x}|)-h(|\vec{k}|,1;t,|\vec{x}|)}{c_*-c}\\
    &+\int_0^\infty |\vec{k}|^2 d|\vec{k}| \, h(|\vec{k}|,1;t,|\vec{x}|) \ln \left(\frac{c_*-c_{\mathrm{cut}}}{c_*-1}\right),
\end{aligned}
\end{equation}
where the angular integration is split into the region $[-1,c_{\text{cut}}]$, which is free of the near-collinear numerical difficulty, and the region $[c_{\text{cut}},1]$, where this difficulty arises. The former contribution is integrated directly. In the latter region, the near-collinear singularity is removed by subtracting $h(|\vec{k}|,1;t,|\vec{x}|)$ from the numerator; the subtracted contribution is then integrated analytically and added back.

In practice, we find that, even after subtracting the near-collinear structure, the denominator still develops a sharp peak for $c\to 1$ and $r_\ell\to 0$ near $|\vec{k}|\sim \frac{m_P}{2}(1-r_\ell)$, due to
\begin{equation}
    \frac{1}{E_\ell}\sim \frac{1}{\sqrt{(|\vec{k}|-\frac{m_P}{2}(1-r_\ell))^2+m_P^2 r_\ell}}.
\end{equation}
We therefore apply the change of variables
\(
    |\vec{k}|=\frac{m_P}{2}(1-r_\ell)+m_P\sqrt{r_\ell}\sinh(u)
\)
to the subtracted integral to improve the stability of the numerical integration. With these techniques, the two-dimensional numerical integration for $m_\ell=m_e$ is accelerated by approximately a factor of $20$--$30$.

\section{Point-like contribution to the virtual correction}
\label{app:pointlike_analytic}
\subsection{Coulomb gauge}
The Coulomb-potential and transverse-photon components of the point-like contribution in Coulomb gauge are given by
\begin{equation}\label{delta_TB_M_Coulomb_pt}
\begin{aligned}
    &\delta_{\text{pt},\text{Coul}}^{\text{vir,pot}}(m_P,m_\ell,m_W)=i\frac{2\pi\alpha}{f_P m_P^2 m_\ell^2(1-r_\ell)}\\
    &\times \int\frac{d^4k}{(2\pi)^4}\frac{-m_W^2}{k^2-m_W^2+i\epsilon}\frac{L_{\mu 0}(k,p,p_\ell)H_{M,\text{pt}}^{\mu 0}(k,p)}{\vec{k}^2\left((p_\ell-k)^2-m_\ell^2+i\epsilon\right)},\\
    &\delta_{\text{pt},\text{Coul}}^{\text{vir,tr}}(m_P,m_\ell,m_W)=i\frac{2\pi\alpha}{f_P m_P^2 m_\ell^2(1-r_\ell)}\\
    &\times \int\frac{d^4k}{(2\pi)^4}\frac{-m_W^2}{k^2-m_W^2+i\epsilon}\\
    &\times\frac{\sum_{\lambda=1}^2 \varepsilon_\sigma(k,\lambda) \varepsilon^*_\rho(k,\lambda)L_{\mu}^\sigma(k,p,p_\ell)H_{M,\text{pt}}^{\mu \rho}(k,p)}{\left(k^2+i\epsilon\right)\left((p_\ell-k)^2-m_\ell^2+i\epsilon\right)}.
\end{aligned}
\end{equation}
Here, the $W$-boson mass is introduced as \textcolor{revision}{a} UV regulator for each component. No IR regulator is required in Coulomb gauge.

A Feynman-parameter integration gives the analytic results for these contributions in the large-$m_W$ limit, with terms suppressed by powers of $1/m_W$ omitted:
\begin{equation}\label{delta_TB_M_Coulomb_pt_analytic}
\begin{aligned}
    &\delta_{\text{pt},\text{Coul}}^{\text{vir,pot}}(m_P,m_\ell,m_W)
    =\frac{\alpha}{\pi}\left[\frac{5-2 r_\ell}{6} \ln \frac{m_W^2}{m_P^2}-\frac{1+r_\ell}{1-r_\ell}\right.\\
    &\times\left(2 \operatorname{Li}_2(-r_\ell)+\operatorname{Li}_2(1-r_\ell)+2 \ln r_\ell \ln (1+r_\ell)+\frac{\pi^2}{6}\right)\\
    &\left.+\frac{6 r_\ell\left(2 r_\ell^2-3 r_\ell+3\right) \ln r_\ell-22 r_\ell^3+63 r_\ell^2-72 r_\ell+31}{18(1-r_\ell)^2}\right],\\
    &\delta_{\text{pt},\text{Coul}}^{\text{vir,tr}}(m_P,m_\ell,m_W)
    =\frac{\alpha}{\pi}\left[-\frac{5-2 r_{\ell}}{6} \ln \frac{m_W^2}{m_P^2}\right.\\
    &\left.-\frac{4 r_\ell^3-3 r_\ell^2+18 r_\ell-3}{6(1-r_\ell)^2} \ln r_\ell-\frac{44 r_\ell^2+17 r_\ell+35}{36(1-r_\ell)} \right].
\end{aligned}
\end{equation}
The total Coulomb-gauge point-like contribution is then
\begin{equation}\label{delta_TB_M_Coulomb_pt_analytic_total}
\begin{aligned}
    &\delta_{\text{pt},\text{Coul}}^{\text{vir}}(m_P,m_\ell)\\
    =&\frac{\alpha}{\pi}\left[\frac{1-4r_\ell-r_\ell^2}{2(1-r_\ell)^2}\ln r_\ell+\frac{11r_\ell^2-14r_\ell+3}{4(1-r_\ell)^2}-\frac{1+r_\ell}{1-r_\ell}\right.\\
    \times&\left.\left(2\,{\rm Li}_2(-r_\ell)+{\rm Li}_2(1-r_\ell)+2\ln r_\ell\ln(1+r_\ell)+\frac{\pi^2}{6}\right)\right].
\end{aligned}
\end{equation}
The absence of \textcolor{revision}{a} UV divergence in this expression is a consequence of using the point-like approximation. In the physical short-distance region, however, this approximation breaks down, and the UV behavior of QCD with quark and gluon degrees of freedom must be taken into account. 
The corresponding integral therefore still contains \textcolor{revision}{a} UV divergence, which is canceled in the difference between the $e$ and $\mu$ channels.

\subsection{Feynman gauge}
Here, we evaluate the point-like contribution to the IR-subtracted finite part, $\delta^{\text{vir},\text{con}}_{\text{pt},\text{Feyn}}(m_P,m_\ell;t_s)$, defined in Eq.~(\ref{calc_Feyn}). Unlike in Coulomb gauge, the virtual-photon loop is IR divergent in Feynman gauge. Our strategy is to introduce a photon mass $m_\gamma$ as an IR regulator and evaluate separately the total point-like contribution $\delta_{\text{pt},\text{Feyn}}^{\text{vir}}$ and the contribution from the IR-divergent hadronic function $H_{M,\text{div}}^{\mu\rho}(k,p;t_s)$ defined in Eq.~(\ref{eq:HMdiv}). We then take their difference and subsequently take the $m_\gamma\to 0$ limit to obtain $\delta^{\text{vir},\text{con}}_{\text{pt},\text{Feyn}}(m_P,m_\ell;t_s)$.

The total point-like contribution is given by
\begin{equation}\label{delta_TB_M_Feynman_pt}
\begin{aligned}
    &\delta_{\text{pt},\text{Feyn}}^{\text{vir}}(m_P,m_\ell,m_\gamma,m_W)\\
    =&-i\frac{2\pi\alpha}{f_P m_P^2 m_\ell^2(1-r_\ell)}\int\frac{d^4k}{(2\pi)^4}\frac{-m_W^2}{k^2-m_W^2+i\epsilon}\\
    &\times\frac{L_{\mu\rho}(k,p,p_\ell)H_{M,\text{pt}}^{\mu\rho}(k,p)}{(k^2-m_\gamma^2+i\epsilon)((p_\ell-k)^2-m_\ell^2+i\epsilon)}.
\end{aligned}
\end{equation}
Here, the small photon mass $m_\gamma$ and the large $W$-boson mass $m_W$ regulate the IR and UV divergences, respectively.

Inserting the explicit form of $H_{M,\text{pt}}^{\mu\rho}(k,p)$ given in Eq.~(\ref{eq:HM_def}), a Feynman-parameter integration yields the analytic result in the large-$m_W$ and small-$m_\gamma$ limits, with terms suppressed by powers of $1/m_W$ or $m_\gamma$ omitted:
\begin{equation}\label{vir_pt_all}
\begin{aligned}
    &\delta_{\text{pt},\text{Feyn}}^{\text{vir}}(m_P,m_\ell,m_\gamma,m_W)\\
    =&\frac{\alpha}{4 \pi}\left[\ln \frac{m_P^2}{m_W^2}+2\frac{1+r_{\ell}}{1-r_\ell} \ln r_{\ell} \ln \frac{m_P^2}{m_\gamma^2}\right.\\
    &\left.+2\frac{1-3r_\ell}{1-r_{\ell}}\ln r_{\ell}+\frac{1+r_{\ell}}{1-r_{\ell}} \ln ^2 r_{\ell}+\frac{7}{2}\right].
\end{aligned}
\end{equation}
Its UV divergence ($\propto\ln m_W^2$) is universal between the $e$ and $\mu$ channels and cancels in their difference. Its IR divergence is removed by subtracting the contribution from $H_{M,\text{div}}^{\mu\rho}(k,p;t_s)$ defined in Eq.~(\ref{eq:HMdiv}):
\begin{equation}\label{vir_pt_div}
\begin{aligned}
    &\delta_{\text{div},\text{Feyn}}^{\text{vir}}(m_P,m_\ell,m_\gamma;t_s)=-i\frac{2\pi\alpha}{f_P m_P^2 m_\ell^2(1-r_\ell)}\\
    &\times\int\frac{d^4k}{(2\pi)^4}\frac{L_{\mu\rho}(k,p,p_\ell)H_{M,\text{div}}^{\mu\rho}(k,p;t_s)}{(k^2-m_\gamma^2+i\epsilon)((p_\ell-k)^2-m_\ell^2+i\epsilon)}.
\end{aligned}
\end{equation}
This term is UV finite and therefore requires no $W$ regulator. It is regulated in the IR region by the same photon mass $m_\gamma$ and is evaluated by first performing the $k^0$ integration analytically, followed by a numerical integration over $\vec{k}$.

Taking the difference between $\delta_{\text{pt},\text{Feyn}}^{\text{vir}}$ and $\delta_{\text{div},\text{Feyn}}^{\text{vir}}$, and evaluating the difference between the $e$ and $\mu$ channels, we obtain the IR-convergent part of the point-like contribution in Feynman gauge. The $\ln m_W^2$ terms cancel between the two lepton channels, and the $m_\gamma\to 0$ limit is taken at the end.

\section{Finite-volume correction}
\label{app:finite_volume_tensor}

\begingroup
\color{revision}
\hypersetup{linkcolor=revision,citecolor=revision,urlcolor=revision}
\subsection{Finite-volume corrections using a ground-state model}
\label{app:single_particle_fv}
Residual finite-volume effects in the IVR method are exponentially suppressed as the spatial extent $L$ increases. Using the ground-state model illustrated in Fig.~\ref{fig:fvorder} and described in Appendix D of Ref.~\cite{Boyle:2025uuh}, we obtain these corrections from the difference between its predictions at the reference volume and at the lattice volume in the limit $L_{\rm ref}\to\infty$.

\begin{figure}
\centering
\includegraphics[width=.95\columnwidth]{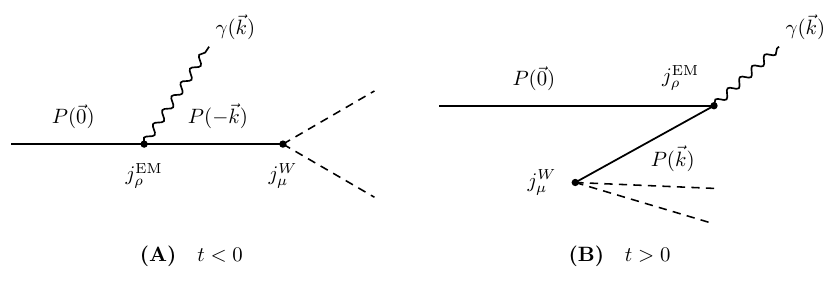}
\captionsetup{labelfont={color=revision},textfont={color=revision}}
\caption{Two time orderings of the ground-state model. Solid, dashed, and wavy lines denote mesons, leptons, and photons, respectively. The diagrams illustrate the hadronic part; the virtual-photon connection to the lepton is omitted.}
\label{fig:fvorder}
\end{figure}

The two time orderings in Fig.~\ref{fig:fvorder} give the following contributions in finite-volume Euclidean space for $t<0$ and $t>0$, respectively:
\begin{multline}
 H^{(L)}_{\mu\rho,A}(t,\vec x)
 =-\frac{1}{L^3}\sum_{\vec k\in\Gamma}
 \frac{q_\mu(q+p_E)_\rho}{2m_PE_P(\vec k)}\\
 \times F^{(P)}\bigl(-(q-p_E)^2\bigr)e^{[E_P(\vec k)-m_P]t+i\vec k\cdot\vec x},
\label{eq:fv_A}
\end{multline}
\begin{multline}
 H^{(L)}_{\mu\rho,B}(t,\vec x)
 \simeq-\frac{1}{L^3}\sum_{\vec k\in\Gamma}
 \frac{q'_\mu(q'-p_E)_\rho}{2m_PE_P(\vec k)}\\
 \times F^{(P)}\bigl(-(q'+p_E)^2\bigr)e^{-[E_P(\vec k)+m_P]t+i\vec k\cdot\vec x}.
\label{eq:fv_B}
\end{multline}
Here $p_E=(im_P,\vec0)$, $q=(iE_P(\vec{k}),-\vec k)$, and $q'=(iE_P(\vec{k}),\vec k)$, with $E_P(\vec k)=\sqrt{m_P^2+\vec k^2}$ and $\Gamma=\{2\pi\vec n/L:\vec n\in\mathbb Z^3\}$. The $t<0$ model contains a single-meson intermediate state; the $t>0$ model retains the $P(\vec0)P(\vec k)$ contribution with the meson at rest as a spectator. The electromagnetic form factor $F^{(P)}(s)$ is evaluated at spacelike and timelike momentum transfers for $t<0$ and $t>0$, respectively. Setting $F^{(P)}(s)=1$ recovers the point-particle contributions in the two time orderings~\cite{Boyle:2025uuh}. Here we retain electromagnetic form factors, which incorporate hadronic structure information such as the meson charge radius. Appendix D of Ref.~\cite{Boyle:2025uuh} gives their parametrizations and shows that the model describes the long-distance lattice data well.

Applying the same IVR formula to the model at volumes $L^3$ and $L_{\rm ref}^3$, then taking $L_{\rm ref}\to\infty$, gives the finite-volume correction
\begin{equation}
 \delta^{\rm FV}_{\rm IVR}=\lim_{L_{\rm ref}\to\infty}\left(\delta_{\rm model}^{(L_{\rm ref})}-\delta_{\rm model}^{(L)}\right).
 \label{eq:fv_correction}
\end{equation}
Here $\delta_{\rm model}^{(L)}$ and $\delta_{\rm model}^{(L_{\rm ref})}$ are the model contributions to the radiative correction at the lattice and reference volumes, respectively. We apply this procedure to real photons and to virtual photons in both gauges.

Taking the pion virtual-photon SD correction $\delta_{\rm SD}^{\rm vir}$ as an example, Fig.~\ref{fig:fvconv} compares the residual finite-volume effects in the two gauges as $L_{\rm ref}$ increases. Using the statistical reference $\sigma_{\rm stat}$ shown in the figure, representative of the Feynman-gauge statistical uncertainty, the residual effects fall below $\sigma_{\rm stat}$ at $L_{\rm ref}\approx9\,\mathrm{fm}$ in Coulomb gauge and below $0.1\sigma_{\rm stat}$ at about $12\,\mathrm{fm}$. The corresponding extents in Feynman gauge are about $21\,\mathrm{fm}$ and $27\,\mathrm{fm}$. The residual effects in both gauges decay at large volumes with the same exponential factor $e^{-m_\pi L_{\rm ref}/2}$, but their smaller magnitude in Coulomb gauge allows a substantially smaller reference volume for the same precision. We use the $L_{\rm ref}\to\infty$ limit for the finite-volume corrections in both gauges.

\begin{figure}
\centering
\includegraphics[width=.92\columnwidth]{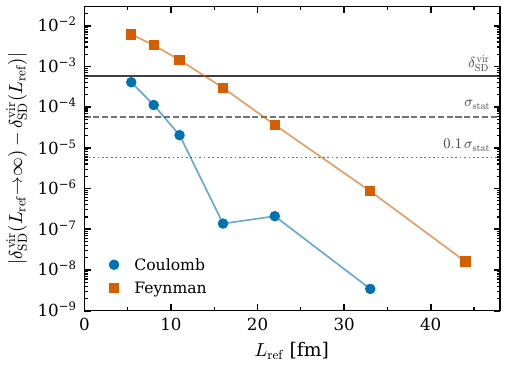}
\captionsetup{labelfont={color=revision},textfont={color=revision}}
\caption{Residual finite-volume effect $|\delta_{\rm SD}^{\rm vir}(L_{\rm ref}\to\infty)-\delta_{\rm SD}^{\rm vir}(L_{\rm ref})|$ in the pion virtual-photon SD correction as a function of $L_{\rm ref}$. Circles and squares denote Coulomb and Feynman gauges, respectively. Each reference length is the mean of the corresponding reference-box extents for 48I and 64I. The solid black line marks the magnitude of the Coulomb-gauge virtual-photon SD correction; the dashed and dotted lines mark the statistical reference $\sigma_{\rm stat}$, representative of the Feynman-gauge statistical uncertainty, and $0.1\sigma_{\rm stat}$, respectively.}
\label{fig:fvconv}
\end{figure}
\endgroup

\subsection{Test of finite-volume effects using 24D and 32D}
We assess the finite-volume correction and estimate the residual finite-volume errors using the 24D and 32D ensembles, which differ only in volume (Table~\ref{table:ens}). To identify the origin of finite-volume effects, we decompose the loop-integral result into contributions associated with different tensor structures of $H_{\mu\nu}(t,\vec{x})$.

\begin{figure*}
    \centering
    \renewcommand{\thesubfigure}{\roman{subfigure}}
    \begin{subfigure}{\textwidth}
        \centering
        \includegraphics[width=0.8\textwidth]{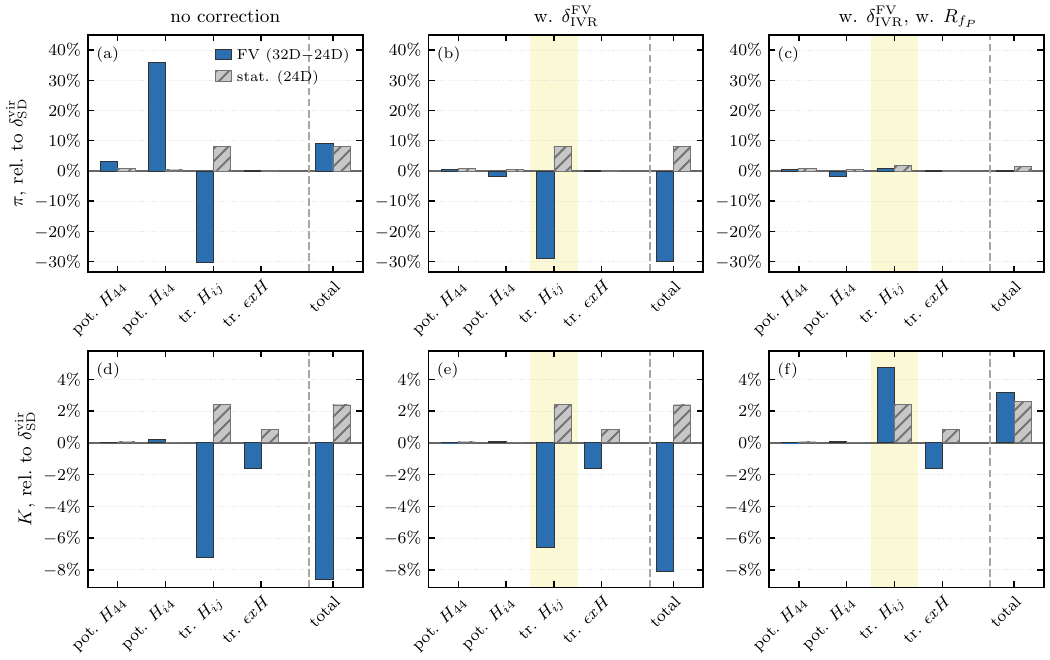}
        \caption{Coulomb gauge. The horizontal axis lists the four channel contributions and their sum: the Coulomb-potential contributions from $H_{44}$ and \textcolor{revision}{$H_{i4}$}, and the transverse-photon contributions from $H_{ij}$ (axial-vector-current part) and $\epsilon xH$ (vector-current part).}
        \label{fig:fv_coul}
    \end{subfigure}\\[0.5ex]
    \begin{subfigure}{\textwidth}
        \centering
        \includegraphics[width=0.8\textwidth]{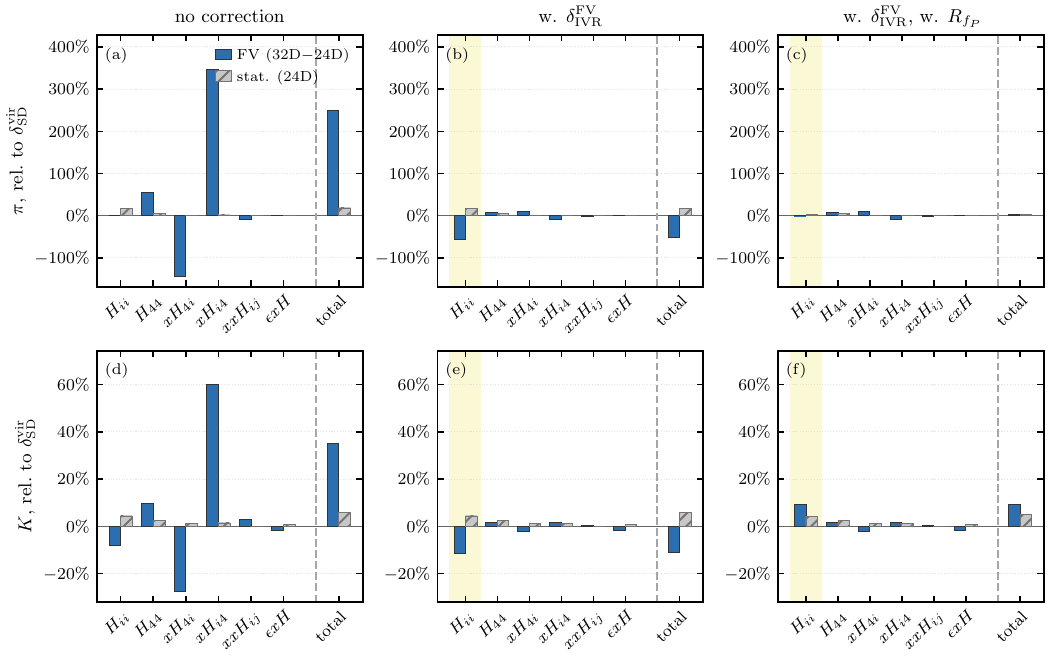}
        \caption{Feynman gauge. The horizontal axis lists the 6 channels and their sum.}
        \label{fig:fv_feyn}
    \end{subfigure}
    \caption{Channel-by-channel decomposition of the finite-volume effects and statistical errors of $\delta_{\text{SD}}^{\text{vir}}$ in (i) Coulomb gauge and (ii) Feynman gauge. In each panel, the blue bars show the finite-volume effect estimated from the $32\text{D}-24\text{D}$ difference, while the gray hatched bars show the statistical error of the 24D result, with both expressed relative to the 24D value of $\delta_{\text{SD}}^{\text{vir}}$. The three columns correspond to successive treatments: no finite-volume correction, with the finite-volume correction, and with the $R_{f_P}$-cancellation technique in addition. The $H_{ii}$ channel \textcolor{revision}{(tr.\ $H_{ij}$ in Coulomb gauge)} is highlighted in pale yellow.}
    \label{fig:fv}
\end{figure*}

Figure~\ref{fig:fv} shows the finite-volume effect in each gauge, estimated channel by channel from the difference between the 24D and 32D results (blue bars), together with the 24D statistical error (gray hatched bars). All data are presented as percentages of the 24D result. The three columns correspond to successive treatments: (i) no finite-volume correction, (ii) inclusion of the finite-volume correction, and (iii) inclusion of both the finite-volume correction and the $R_{f_P}$-cancellation technique. The channels are labeled by the tensor structures of $H_{\mu\nu}(t,\vec{x})$. In Feynman gauge, these are abbreviated as $H_{ii}\equiv H_{ii}(t,\vec{x})$, $H_{44}\equiv H_{44}(t,\vec{x})$, $xH_{4i}\equiv x^i H_{4i}(t,\vec{x})$, $xH_{i4}\equiv x^i H_{i4}(t,\vec{x})$, $xxH_{ij}\equiv x^i x^j H_{ij}(t,\vec{x})$, and $\epsilon x H\equiv \varepsilon^{ijk}x_k H_{ij}(t,\vec{x})$, with repeated spatial indices $i,j\in\{1,2,3\}$ summed. In Coulomb gauge, the Coulomb-potential part receives contributions from $H_{44}(t,\vec{x})$ and \textcolor{revision}{$H_{i4}(t,\vec{x})$}, labeled pot.\ $H_{44}$ and \textcolor{revision}{pot.\ $H_{i4}$}, respectively. The transverse-photon part receives contributions from the axial-vector current, through $H_{ij}(t,\vec{x})$ and labeled tr. $H_{ij}$, and from the vector current, through $\varepsilon^{ijk}x_k H_{ij}(t,\vec{x})$ and labeled tr. $\epsilon x H$. The latter coincides with its Feynman-gauge counterpart.

Figure~\ref{fig:fv} shows that the choice of gauge has a strong impact on the size of finite-volume effects. In the uncorrected case (left columns), the finite-volume effect in Feynman gauge reaches $250\%$ of the signal for $\pi$ and $35\%$ for $K$, dominated by the $xH_{4i}$ and $xH_{i4}$ channels. In Coulomb gauge, by contrast, it is reduced to only $9\%$ for $\pi$ and $K$. This substantial reduction arises because Coulomb gauge avoids the IR divergence, leading to a much smaller point-like contribution and hence a much smaller finite-volume effect associated with it.

The correction based on the ground-state model (middle column) removes most of the finite-volume effect in the $H_{44}$, $H_{4i}$, and $H_{i4}$ channels. As discussed in Ref.~\cite{Boyle:2025uuh}, this correction accounts for the dominant finite-volume effect associated with the point-like contribution. The residual effects are then concentrated in the $H_{ii}$ channel, which receives only a small contribution from the single-particle state and is instead dominated by heavier $J^P=1^-$ intermediate states, including $\pi\pi$, $K\pi$, and the vector mesons $\rho$ and $K^*$. The FV effects from these contributions cannot be captured by the ground-state model. This pattern is observed in both gauges. In Feynman gauge, this correction removes the dominant finite-volume effect but leaves a residual contribution in the $H_{ii}$ channel, which for $\pi$ still reaches $O(50\%)$ of the signal. In Coulomb gauge, the finite-volume effect is already strongly suppressed relative to that in Feynman gauge; after applying the correction based on the ground-state model, the residual is dominated by the transverse-photon contribution, in particular its $H_{ii}$ component.

Figure~\ref{fig:fv} shows that both the residual finite-volume effect and the statistical error are concentrated in the $H_{ii}$ channel, as highlighted by the yellow bands. The $R_{f_P}$-cancellation technique introduced in the main text is designed to target precisely the errors from this contribution. The quantity $R_{f_P}$ defined in Eq.~(\ref{eq:RfP}) encodes the zero-momentum projection of the $H_{ii}$ matrix element. For $\pi$, the small pion mass makes the $H_{ii}$ contribution to the loop integral strongly correlated with $R_{f_P}$; dividing by $R_{f_P}$ (third column) therefore cancels much of both the finite-volume effect and the statistical error in both gauges. For the kaon, this correlation is weaker, and the technique does not yield a comparable reduction in statistical uncertainty. Tables~\ref{tab:M2full} and \ref{tab:M2} list the results with and without this technique.

\section{Real-photon emission}
\label{app:real_photon}
This section describes the computation of the SD correction from $O(\alpha)$ real-photon emission, $\delta_{\text{SD}}^{\text{real}}(m_P,m_\ell)$. We compute it on the lattice for both $\pi$ and $K$; under the convention specified below Eq.~\eqref{Remu_def1}, only the pion result enters $R_{e/\mu}$.

This correction can be written as an integral over the three-body phase space of the SD part of the reduced squared amplitude for the radiative decay,
\begin{equation}\label{eq:real_def}
    \delta_{\text{SD}}^{\text{real}}(m_P,m_\ell)=\frac{\alpha}{2\pi(1-r_\ell)^2}\int_{\Omega_3} d x_\gamma\, d y_\ell\, A_{\text{SD}}(x_\gamma,y_\ell),
\end{equation}
where the three-body phase-space variables $(x_\gamma,y_\ell)$ and integration region $\Omega_3$ are defined by
\begin{equation}\label{eq:real_phasespace}
\begin{aligned}
    &x_\gamma=\frac{2 p \cdot k}{m_P^2},\quad y_{\ell}=\frac{2 p \cdot p_{\ell}}{m_P^2},\\
    &0 \leq x_\gamma \leq 1-r_{\ell},\quad 1-x_\gamma+\frac{r_{\ell}}{1-x_\gamma} \leq y_{\ell} \leq 1+r_{\ell}.
\end{aligned}
\end{equation}
The reduced squared amplitude $A(x_\gamma,y_\ell)$ is decomposed in the conventional form
\begin{equation}\label{eq:real_amp}
\begin{aligned}
    A(x_\gamma,y_\ell)=&\,f_{\mathrm{IB}}(x_\gamma,y_\ell)\\
    +&\frac{1}{r_{\ell}}\Big(\frac{m_P}{2 f_P}\Big)^2\Big[(F_V+F_A)^2 f_{\mathrm{SD}^{+}}(x_\gamma,y_\ell)\\
    +&(F_V-F_A)^2 f_{\mathrm{SD}^{-}}(x_\gamma,y_\ell)\Big]\\
    -&\frac{m_P}{f_P}\Big[(F_V+F_A) f_{\mathrm{INT}^{+}}(x_\gamma,y_\ell)\\
    +&(F_V-F_A) f_{\mathrm{INT}^{-}}(x_\gamma,y_\ell)\Big],\\
    A_{\text{SD}}(x_\gamma,y_\ell)=&\,A(x_\gamma,y_\ell)-f_{\mathrm{IB}}(x_\gamma,y_\ell).
\end{aligned}
\end{equation}
The kinematic functions $f_{\mathrm{IB}}(x_\gamma,y_\ell)$, $f_{\mathrm{SD}^{\pm}}(x_\gamma,y_\ell)$, and $f_{\mathrm{INT}^{\pm}}(x_\gamma,y_\ell)$ are given in Ref.~\cite{Boyle:2025uuh}. The inner-bremsstrahlung term $f_{\mathrm{IB}}(x_\gamma,y_\ell)$ is already included in the point-like contribution $\Delta_{\alpha,\text{pt}}$. The SD part $A_{\text{SD}}(x_\gamma,y_\ell)$, consisting of the form-factor-squared ($\mathrm{SD}^{\pm}$) and form-factor--point-like interference ($\mathrm{INT}^{\pm}$) terms, is IR finite and depends on the vector and axial-vector form factors $F_V(x_\gamma)$ and $F_A(x_\gamma)$. In Ref.~\cite{Boyle:2025uuh}, we determined these form factors by IVR on the same 48I and 64I ensembles used here. \textcolor{revision}{We use that method to compute the form factors and then evaluate $\delta_{\text{SD}}^{\text{real}}(m_P,m_\ell)$ using Eq.~(\ref{eq:real_def}).}

\section{Meson-mass correction}
\label{app:meson_mass_correction}

We use the 48I and 64I ensembles listed in Table~\ref{table:ens}, which have similar physical volumes but different lattice spacings, for the continuum extrapolation. However, the kaon masses on the two ensembles differ slightly: $m_K^{\text{48I}}\simeq 499.2~\text{MeV}$ and $m_K^{\text{64I}}\simeq 508.0~\text{MeV}$. A direct $a^2$-linear continuum extrapolation of $\delta_{\text{SD}}^{\text{vir}}(m_K,m_\ell)$ would therefore introduce a bias from this kaon-mass mismatch. To avoid this, we correct the 64I result to the 48I kaon mass before performing the continuum extrapolation.

We compute partially quenched propagators and correlators on the same 64I ensemble to determine the dependence of $\delta_{\text{SD}}^{\text{vir}}(m_K,m_\ell)$ on $m_K^2$. We neglect effects from the mismatch between sea- and valence-quark masses, which enter through disconnected diagrams. 
For the closely spaced kaon masses used in 48I and 64I, we assume that the observable depends linearly on $m_K^2$.
The kaon-mass correction is performed using the 64I-pq2 setup in Table~\ref{table:ens}, which has a lighter valence strange-quark mass and the same light-quark mass as 64I. This setup therefore isolates the dependence on $m_K^2$.
Using the same $31$ configurations for the 64I and 64I-pq2 setups, we determine the slope of the observable with respect to the squared kaon mass,
\begin{equation}\label{eq:mass_slope}
    \frac{\partial\delta_{\text{SD}}^{\text{vir}}}{\partial m_K^2}
      \;=\; \frac{\delta_{\text{SD}}^{\text{vir}}(m_K^{\text{64I-pq2}},m_\ell)
                 -\delta_{\text{SD}}^{\text{vir}}(m_K^{\text{64I}},m_\ell)}
                 {\left(m_K^{\text{64I-pq2}}\right)^2-\left(m_K^{\text{64I}}\right)^2}.
\end{equation}
Using this slope, we implement the kaon-mass correction in three steps:
\begin{enumerate}
    \item We first correct the 64I result from $m_K^{\text{64I}}$ to $m_K^{\text{48I}}$,
    \begin{equation}
    \begin{aligned}
        \delta_{\text{SD}}^{\text{vir}}\!\big|_{m_K^{\text{64I}} \to m_K^{\text{48I}}}
        &= \delta_{\text{SD}}^{\text{vir}}(m_K^{\text{64I}},m_\ell)\\
        &+ \frac{\partial\delta_{\text{SD}}^{\text{vir}}}{\partial m_K^2}\left[\left(m_K^{\text{48I}}\right)^2-\left(m_K^{\text{64I}}\right)^2\right].
        \label{eq:corr64}
    \end{aligned}
    \end{equation}
    \item We then perform the $a^2$-linear continuum extrapolation using the corrected 64I result together with the 48I result, obtaining the continuum-extrapolated result $\delta_{\text{SD}}^{\text{vir,cont}}(m_K^{\text{48I}},m_\ell)$ at $m_K^{\text{48I}}=499.2~\text{MeV}$.
    \item Finally, we shift the continuum-extrapolated result to the physical charged-kaon mass, $m_{K^\pm}=493.677~\text{MeV}$, using the same slope,
    \begin{equation}
    \begin{aligned}
        \delta_{\text{SD}}^{\text{vir,cont}}(m_{K^\pm},m_\ell)
        &= \delta_{\text{SD}}^{\text{vir,cont}}(m_K^{\text{48I}},m_\ell)\\
        &+ \frac{\partial\delta_{\text{SD}}^{\text{vir}}}{\partial m_K^2}\left[\,m_{K^\pm}^2-\left(m_K^{\text{48I}}\right)^2\right].
        \label{eq:shift_phys}
    \end{aligned}
    \end{equation}
    This step neglects the lattice-spacing dependence of the slope; since the shift itself is smaller than the statistical uncertainty, the error associated with this approximation is negligible at the current precision.
\end{enumerate}
\textcolor{revision}{The mass slope is determined from 31 configurations that form a subset of the 119 used for the main 64I calculation. We use a stratified version of the bootstrap method~\cite{Davison_1997}. For each bootstrap sample, we resample the subset of 31 configurations and the remaining 88 configurations separately. We use the same resampling indices for the 31-configuration subset at both mass points to determine the mass slope. At the unitary mass, the resampled data from this subset are combined with those from the remaining 88 configurations to obtain the main result. The slope is then applied to the main result within the same bootstrap sample according to Eq.~\eqref{eq:corr64}, and the statistical uncertainty is estimated from the standard deviation of the mass-corrected samples. Since the main result and the mass slope use exactly the same resampling indices for the 31 configurations, their statistical correlation is preserved.}

\begin{figure}
    \centering
    \includegraphics[width=\columnwidth]{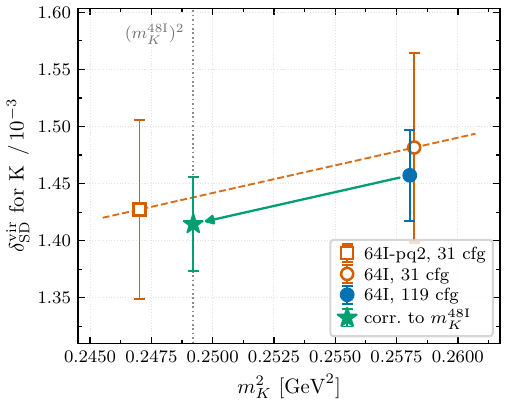}
    \caption{Kaon-mass correction to $\delta_{\text{SD}}^{\text{vir}}$ for the kaon on the 64I ensemble. The filled circle denotes the result from all $119$ configurations of 64I. The open circle and open square denote the 64I and 64I-pq2 results, respectively, computed on the same $31$ configurations and used to determine the slope of $\delta_{\text{SD}}^{\text{vir}}$ with respect to $m_K^2$ (orange dashed line). Along this slope, the 64I result from all $119$ configurations is corrected to the 48I kaon mass, $(m_K^{\text{48I}})^2$ (vertical dotted line), giving the corrected value shown as the star.}
    \label{fig:mass_corr}
\end{figure}

Figure~\ref{fig:mass_corr} illustrates this procedure using the Coulomb-gauge result as an example. Since $m_K^{\text{48I}}$ lies between $m_K^{\text{64I-pq2}}$ and $m_K^{\text{64I}}$, correcting the 64I result to $(m_K^{\text{48I}})^2$ amounts to a linear interpolation between the two points. The Feynman-gauge result is obtained with the identical procedure.

Finally, we verify that the pion-mass correction is negligible. Although $\delta_{\text{SD}}$ does depend on $m_\pi^2$, the pion masses on 48I ($139.55~\text{MeV}$) and 64I ($139.18~\text{MeV}$) already lie very close to the physical value $m_{\pi^\pm}=139.57~\text{MeV}$, so the correction over this small gap is tiny. Using the 64I-pq1 ensemble, the $m_\pi^2$ correction shifts the pion continuum result by only about $0.01\sigma$ of the statistical error. For the kaon, a two-dimensional linear fit in $(m_\pi^2,m_K^2)$ using 64I, 64I-pq1, and 64I-pq2 gives an $m_\pi$-correction effect of about $0.02\sigma$. Both are far below the current statistical error, so only the kaon-mass correction is applied.

\section{Higher-order QED uncertainty}
\label{app:higher_order_qed}
\subsection{Estimate of the NLL contribution to $\Delta_{\alpha^{n\geq 2}}$}
Here, we estimate the size of the NLL contribution to $\Delta_{\alpha^{n\geq 2}}$ using two-loop QED renormalization-group running, and include this estimate in the error budget of our final result.

Ref.~\cite{Marciano:1993sh} gives the LL contribution, namely the terms proportional to $\alpha^n\log^n(m_\mu/m_e)$ with $n\geq 2$. This is obtained by the renormalization-group running of the lepton mass in QED from the scale $\mu=m_\mu$ to $\mu=m_e$. More specifically, at one loop in QED, the running of the fine-structure constant and of a single lepton mass between two scales $\mu_1$ and $\mu_2$ is given by~\cite{Grozin:2005yg}
\begin{equation}
\begin{aligned}
\alpha\left(\mu_2\right) & =\frac{\alpha\left(\mu_1\right)}{1+2\beta_0 L\frac{\alpha\left(\mu_1\right)}{4\pi} }, \\
m\left(\mu_2\right) & =m\left(\mu_1\right)\left[\frac{\alpha\left(\mu_2\right)}{\alpha\left(\mu_1\right)}\right]^{\gamma_{m0}/(2\beta_0)},
\end{aligned}
\end{equation}
where $L=\log(\mu_2/\mu_1)$. The expansions of the QED $\beta$ function and mass anomalous dimension are given by
\begin{equation}
\begin{aligned}
    \beta(\alpha)&=\beta_0 \frac{\alpha}{4\pi}+\beta_1 \frac{\alpha^2}{(4\pi)^2}+\cdots,\\
    \gamma_m(\alpha)&=\gamma_{m0} \frac{\alpha}{4\pi}+\gamma_{m1} \frac{\alpha^2}{(4\pi)^2}+\cdots,
\end{aligned}
\end{equation}
where $\beta_0,\beta_1$ and $\gamma_{m0},\gamma_{m1}$ are the one- and two-loop coefficients. Only $\beta_0$ and $\gamma_{m0}$ enter at LL order. Between the scales $m_e$ and $m_\mu$, only the electron is active, so the running is that of single-flavor QED, with $\beta_0=-\frac{4}{3}$ and $\gamma_{m0}=6$.

Since the leptonic decay width is proportional to $m_\ell^2$, the effect on $R_{e/\mu}$ of running the scale from $\mu_2=m_\mu$ to $\mu_1=m_e$ is
\begin{equation}
\begin{aligned}
\Delta_{LL}&=\left[1+2\beta_0\frac{\alpha}{4\pi} L\right]^{-\gamma_{m0}/\beta_0}\\
&=\left[1-\frac{2 \alpha}{3 \pi} L\right]^{9/2}.
\end{aligned}
\end{equation}
This expression resums the leading logarithms from first order in $\alpha$ to all orders. Since the first-order term is already included in $\Delta_{\alpha,\text{pt}}$, it must be subtracted:
\begin{equation}
    \Delta_{\alpha^{n\geq 2},\text{LL}}=\frac{\left[1-\frac{2 \alpha}{3 \pi} L\right]^{9 / 2}}{1-3\frac{ \alpha}{ \pi} L}-1\approx 0.055\%.
\end{equation}
This LL contribution is included in the central value of $R_{e/\mu}$ in the main text, while the NLL and non-logarithmic $O(\alpha^2)$ terms estimated below are taken as the associated theoretical uncertainty.

The above procedure extends naturally to two loops and provides an estimate of the size of the NLL contribution. The two-loop QED renormalization-group running is~\cite{Grozin:2005yg}
\begin{equation}
\begin{aligned}
    \alpha(\mu_2)&=\alpha(\mu_1)\Big[1-2\beta_0 L\frac{\alpha(\mu_1)}{4\pi}\\
    &+\left(4\beta_0^2 L^2-2\beta_1 L\right) \frac{\alpha^2(\mu_1)}{(4\pi)^2}+\cdots\Big],\\
    m(\mu_2)&=m(\mu_1)\left[\frac{\alpha(\mu_2)}{\alpha(\mu_1)}\right]^{\gamma_{m0}/(2\beta_0)}\\
    &\times\bigg[1+\left(\frac{\gamma_{m1}}{2\beta_0}-\frac{\gamma_{m0}\beta_1}{2\beta_0^2}\right)\frac{\alpha(\mu_2)-\alpha(\mu_1)}{4\pi}+\cdots\bigg].
\end{aligned}
\end{equation}
For single-flavor QED, the two-loop coefficients are $\beta_1=-4$ and $\gamma_{m1}=-\frac{11}{3}$. We estimate the size of \textcolor{revision}{the} NLL contribution from the change in $\Delta_{\alpha^{n\geq 2}}$ when the one-loop running is replaced by the two-loop running, which gives $0.0017\%$.

\subsection{Estimate of other $O(\alpha^2)$ contributions}
For the $O(\alpha^2)$ contributions that are not enhanced by $\log(m_\mu/m_e)$, we estimate their size as
\begin{equation}
    [1\sim 5]\times \frac{\alpha^2}{\pi^2}=[0.0005\%\sim 0.0027\%].
\end{equation}
\textcolor{revision}{This contribution is similar in size to the NLL estimate.} We take the more conservative value, $0.0027\%$, as the estimate of these other $O(\alpha^2)$ contributions. Adding this in quadrature with the $0.0017\%$ NLL estimate from the previous subsection gives a total uncertainty of $0.003\%$ from $O(\alpha^2)$ terms beyond the leading logarithm in the final error budget. This uncertainty is still smaller than the current statistical precision of the lattice calculation; however, as lattice precision improves in the future, it will become important.

\begingroup
\color{revision}
\hypersetup{linkcolor=revision,citecolor=revision,urlcolor=revision}
\section{Subtraction of around-the-world effects}
\label{app:atw_subtraction}
In this section, we describe our treatment of around-the-world (ATW) effects arising from periodic temporal boundary conditions. We illustrate the procedure using the pion three-point function
\begin{equation}\begin{aligned}
 C_{\mu\rho}(t,\vec x)=\langle T\{j^W_\mu(0)j^{\rm EM}_\rho(t,\vec x)
 \phi_\pi^\dagger(t_\pi)\}\rangle,\\
 t_\pi=\min(0,t)-\Delta T,
\label{eq:atw_C}
\end{aligned}\end{equation}
where $\Delta T$ is the temporal separation between the wall source and the nearer current.

Figure~\ref{fig:atw_paths} illustrates the signal and ATW contributions for both time orderings. The first column shows the signal without ATW effects, denoted by $C_{\mu\rho}^{\rm sig}$; only the lightest intermediate states are drawn. The second column shows the backward propagation of the signal in the opposite time ordering, denoted by $C_{\mu\rho}^{\rm ATW\text{-}1}$, which is the dominant ATW contribution and can be reconstructed and subtracted using lattice data. The third column shows other ATW contributions, denoted by $C_{\mu\rho}^{\rm ATW\text{-}2}$. Within the chosen plateau window, their effect on the Coulomb-gauge result is negligible at the current statistical precision, as shown below.

\begin{figure}
\centering
\includegraphics[width=\columnwidth]{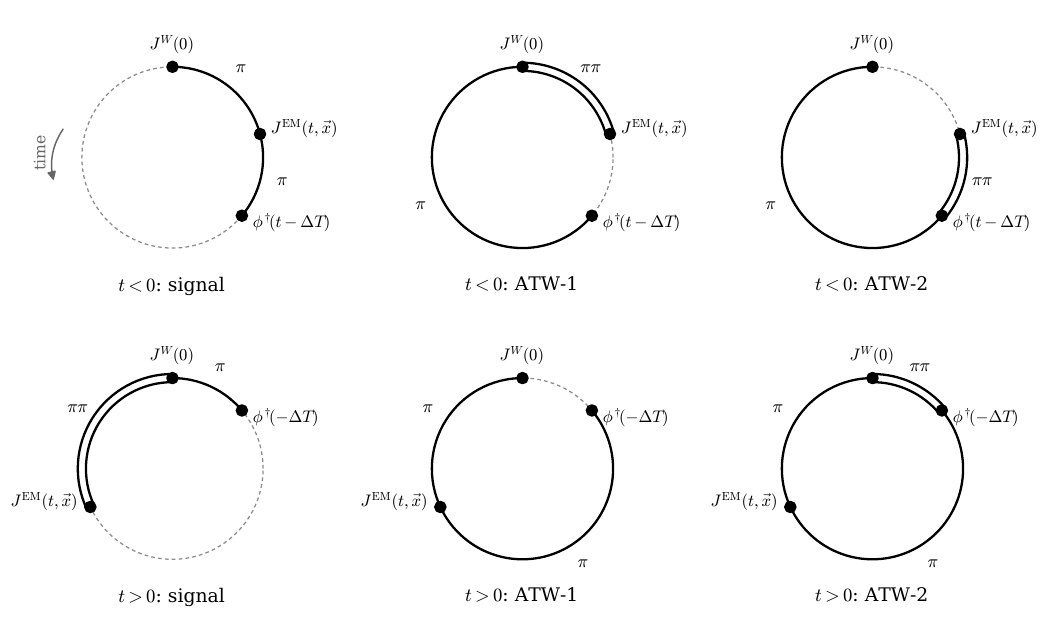}
\captionsetup{labelfont={color=revision},textfont={color=revision}}
\caption{Signal and ATW contributions for $t<0$ (top) and $t>0$ (bottom), showing only the lightest intermediate states. Single, double, and dashed lines denote a pion, a two-pion state, and the vacuum, respectively.}
\label{fig:atw_paths}
\end{figure}

\begin{figure}
\centering
\includegraphics[width=.92\columnwidth]{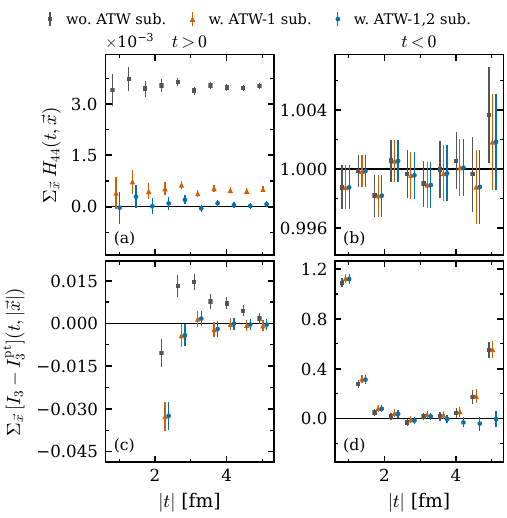}
\captionsetup{labelfont={color=revision},textfont={color=revision}}
\caption{Effect of ATW subtraction on pion hadronic matrix elements on 48I. Left and right columns correspond to $t>0$ and $t<0$, respectively. Upper panels: $\sum_{\vec x}H_{44}(t,\vec{x})$, with horizontal lines marking the charge-conservation values. Lower panels: $\sum_{\vec x}(I_3-I_3^{\rm pt})(t,|\vec{x}|)$, with horizontal lines at zero.}
\label{fig:atw_me}
\end{figure}

To show the relation between ATW-1 and the signal in the opposite time ordering, we retain the pion ground state for propagation from the wall source, while summing over all intermediate states between the currents, labeled by $n$ and $n'$ for $t<0$ and $t>0$, respectively:
\begin{equation}
\begin{aligned}
 C^{\rm sig}_{\mu\rho}(t,\vec x)
 &=\frac{e^{-m_\pi\Delta T}}{L^3}\\
 &\quad\times
 \begin{cases}
 \displaystyle\sum_{\vec k,n}\mathcal A_{\mu\rho,n}(\vec k)
 e^{-E_n(\vec k)|t|-i\vec k\cdot\vec x},&t<0,\\[3pt]
 \displaystyle\sum_{\vec k,n'}\mathcal B_{\mu\rho,n'}(\vec k)
 e^{-E_{n'}(\vec k)|t|+i\vec k\cdot\vec x},&t>0,
 \end{cases}\\[6pt]
 C^{\rm ATW\text{-}1}_{\mu\rho}(t,\vec x)
 &=\frac{e^{-m_\pi(T-|t|-\Delta T)}}{L^3}\\
 &\quad\times
 \begin{cases}
 \displaystyle\sum_{\vec k,n'}\mathcal B_{\mu\rho,n'}(\vec k)
 e^{-E_{n'}(\vec k)|t|-i\vec k\cdot\vec x},&t<0,\\[3pt]
 \displaystyle\sum_{\vec k,n}\mathcal A_{\mu\rho,n}(\vec k)
 e^{-E_n(\vec k)|t|+i\vec k\cdot\vec x},&t>0.
 \end{cases}
\end{aligned}
\label{eq:atw_spectral}
\end{equation}
\begin{figure}
\centering
\includegraphics[width=.92\columnwidth]{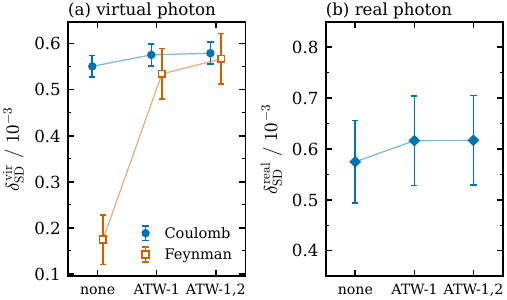}
\captionsetup{labelfont={color=revision},textfont={color=revision}}
\caption{Continuum-extrapolated pion SD corrections with no ATW subtraction, ATW-1 subtraction, and ATW-1 plus ATW-2 subtraction (left to right). Panel (a): virtual-photon corrections in Coulomb and Feynman gauges. Panel (b): real-photon correction.}
\label{fig:atw_switch}
\end{figure}

Here $L$ and $T$ are the spatial and temporal extents, and $\vec k$ runs over the discrete finite-volume momenta. The energies of the intermediate states are $E_n(\vec k)$ and $E_{n'}(\vec k)$, while $\mathcal A_{\mu\rho,n}(\vec k)$ and $\mathcal B_{\mu\rho,n'}(\vec k)$ collect the products of matrix elements and state-normalization factors. Since ATW-1 and the signal in the opposite time ordering share the same spectral coefficients, we obtain, for either sign of $t$,
\begin{equation}
 C^{\rm ATW\text{-}1}_{\mu\rho}(t,\vec x)
 =e^{-m_\pi(T-|t|-2\Delta T)}C^{\rm sig}_{\mu\rho}(-t,-\vec x).
\label{eq:atw_relation}
\end{equation}
This relation allows us to reconstruct ATW-1 directly from lattice data in the opposite time ordering. Since the $t>0$ data are more strongly contaminated by ATW effects than the $t<0$ data, the subtraction can proceed by first using the $t<0$ data to reconstruct and subtract ATW-1 at $t>0$, then using the corrected $t>0$ data to subtract ATW-1 at $t<0$. We have checked that ATW contamination in the $t<0$ input data in the first step has a negligible effect at the current statistical precision.

We estimate ATW-2 using the ground-state model illustrated in Fig.~\ref{fig:fvorder} of Sec.~\ref{app:single_particle_fv}. This model is appropriate because the intermediate states in the ATW-2 diagrams of Fig.~\ref{fig:atw_paths} propagate over long times. We obtain
\begin{equation}\begin{aligned}
 &C^{\mathrm{ATW\text{-}2}}_{\mu\rho}(t,\vec x)\simeq N_\pi\,im_\pi f_\pi\\
 &\quad\times
 \begin{cases}
 e^{-m_\pi(T+\Delta T)}H^{(L)}_{\mu\rho,A}(-(T-t),\vec x),&t>0,\\
 e^{m_\pi(T-|t|-\Delta T)}H^{(L)}_{\mu\rho,B}(T-|t|,\vec x),&t<0.
 \end{cases}
 \label{eq:atw_second}
\end{aligned}\end{equation}
where $H^{(L)}_{\mu\rho,A}$ and $H^{(L)}_{\mu\rho,B}$ are defined in Eqs.~\eqref{eq:fv_A} and~\eqref{eq:fv_B}. The normalization factor $N_\pi=|\langle0|\phi_\pi(0)|\pi\rangle|/(2m_\pi)$, pion mass $m_\pi$, and decay constant $f_\pi$ are determined from two-point functions. 

Figure~\ref{fig:atw_me} tests charge conservation and the long-time behavior of the hadronic matrix elements on 48I before and after ATW subtraction. The upper panels show $\sum_{\vec x}H_{44}(t,\vec x)$; with the normalization in Eq.~\eqref{IE}, charge conservation fixes this sum to $1$ for $t<0$ and $0$ for $t>0$. In panel (a), the ATW contribution is only about $0.3\%$ but remains clearly visible because the signal vanishes. Subtracting ATW-1 and ATW-2 brings the data into agreement with zero within statistical errors. The lower panels show $\sum_{\vec x}(I_3-I_3^{\rm pt})(t,|\vec{x}|)$, where $I_3(t,|\vec{x}|)$ is defined in Eq.~\eqref{eq:scalar_lattice_inputs} and $I_3^{\rm pt}(t,|\vec{x}|)$ is the corresponding contribution from the ground-state model illustrated in Fig.~\ref{fig:fvorder} of Sec.~\ref{app:single_particle_fv}. After ATW subtraction, this difference is consistent with zero at large $|t|$, restoring agreement with the long-time behavior described by the model. In panel (d), ATW-2 grows exponentially with $|t|$ near the temporal boundary. This behavior arises from the factor $e^{-E_\pi(\vec k)(T-|t|)+m_\pi|t|}$ in each momentum component of its contribution to $H_{\mu\rho}(t,\vec{x})$, where $E_\pi(\vec k)$ is the pion energy. For the pion virtual- and real-photon corrections on 48I and 64I, we therefore choose $t_s\in[3.0,3.6]\,\mathrm{fm}$ to avoid this region; see Table~\ref{tab:timecuts}.

Figure~\ref{fig:atw_switch} shows how ATW subtraction affects the continuum-extrapolated pion SD corrections. Within the chosen time windows, ATW-1 shifts the Coulomb-gauge virtual-photon result [Fig.~\ref{fig:atw_switch}(a)] by an amount comparable to its statistical uncertainty, while ATW-2 is negligible at the current precision. By contrast, the figure shows that the Feynman-gauge result is more sensitive to ATW effects. This is because the infrared subtraction in the $H_{44}(t,\vec{x})$ channel amplifies the small ATW-induced violation of charge conservation at $t>0$ [Fig.~\ref{fig:atw_me}(a)] relative to the SD signal.

\endgroup

\begingroup
\color{revision}
\hypersetup{linkcolor=revision,citecolor=revision,urlcolor=revision}
\section{Selection of time-integration limits and source--current separation}
\label{app:timecuts}
In the lattice calculation, we truncate the time integral in Eq.~\eqref{eq:delta_coul_lat}, giving
\begin{equation}\label{eq:timecuts}
\begin{aligned}
    \delta^{\text{vir}}_{\text{Coul}}(m_P,m_\ell)&=\int d^3x\!\int_{-t_s}^{t_{\max}}\!dt\,H_{\mu\rho}(t,\vec{x})\,\mathcal{K}_{\mu\rho}(t,\vec x)\\
    &+\int d^3x~H_{\mu\rho}(-t_s,\vec{x}) \,\mathcal{K}^{(l)}_{\mu\rho}(-t_s,\vec x).
\end{aligned}
\end{equation}
Here $t_{\min}=-t_s<0$ and $t_{\max}>0$ are the temporal cutoffs of the short-distance integral. When IVR is used, $t=-t_s$ separates short-distance integration from long-distance reconstruction. For $R_{f_P}$ and the real-photon contribution, which require no long-distance reconstruction, we also denote the lower integration limit by $-t_s$ for convenience.

Figure~\ref{fig:timecuts} shows the virtual- and real-photon SD corrections and $R_{f_P}$ as functions of $t_s=|t_{\min}|$ and $t_{\max}$ on 48I and 64I. When one endpoint is scanned, the other is restricted to its plateau window. As the integration range increases, the results reach plateaus, indicating that the integrated contributions have converged within statistical errors. We select the windows in Table~\ref{tab:timecuts} from these plateaus, avoiding the region of enhanced negative-time ATW effects discussed in Sec.~\ref{app:atw_subtraction}. The bands show the results obtained from the selected plateau windows.

The pion correction $\delta_{\rm SD,Feyn}^{\rm vir}$ reaches a plateau only at larger $t_{\max}$. This slow convergence arises from the point-particle contribution represented by diagram B in Fig.~\ref{fig:fvorder}. The associated truncation effect is amplified by the large cancellation in the point-particle subtraction. In Coulomb gauge, this contribution is strongly suppressed and the plateau is reached earlier.

\begin{table}[t]
\color{revision}
\captionsetup{labelfont={color=revision},textfont={color=revision}}
\centering\small\setlength{\tabcolsep}{3pt}
\caption{Time-integration windows, in fm. The kaon virtual-photon window applies to both gauges.}\label{tab:timecuts}
\begin{tabular}{lcc}\toprule
Quantity & $t_s=|t_{\min}|$ & $t_{\max}$\\\midrule
$\pi$: virtual (Coul.), real & $[3.0,3.6]$ & $[3.0,4.9]$\\
$\pi$: virtual (Feyn.) & $[3.0,3.6]$ & $[4.5,4.9]$\\
$K$: virtual, real & $[2.0,3.0]$ & $[2.7,3.5]$\\
$\pi$: $R_{f_P}$ & $[3.0,3.8]$ & $[2.7,4.5]$\\
$K$: $R_{f_P}$ & $[2.4,3.0]$ & $[2.5,4.0]$\\\bottomrule
\end{tabular}
\end{table}

\begin{figure*}
\centering
\includegraphics[width=.9\textwidth]{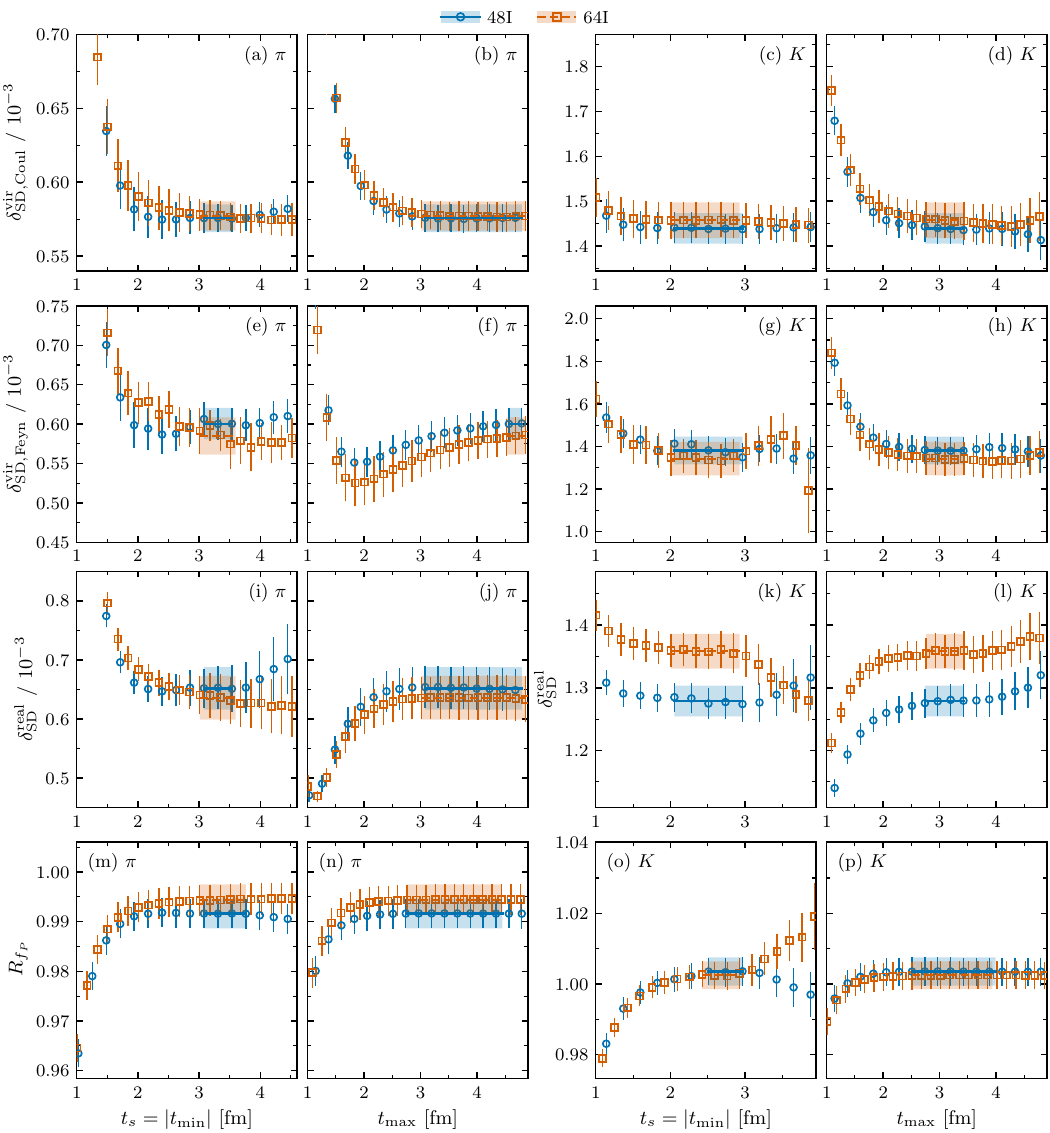}
\captionsetup{labelfont={color=revision},textfont={color=revision}}
\caption{Dependence of the virtual- and real-photon SD corrections and $R_{f_P}$ on the temporal integration limits. Rows show $\delta^{\rm vir}_{\rm SD,Coul}$, $\delta^{\rm vir}_{\rm SD,Feyn}$, $\delta^{\rm real}_{\rm SD}$, and $R_{f_P}$; columns show the $t_s=|t_{\min}|$ and $t_{\max}$ scans for $\pi$ (left pair) and $K$ (right pair). Circles and squares denote 48I and 64I, respectively. Each fit band spans the window in Table~\ref{tab:timecuts}.}
\label{fig:timecuts}
\end{figure*}

The three-point functions are computed with a fixed source--current separation $\Delta T$. With the currents at times $0$ and $t$, the wall-source creation operator $\phi_P^\dagger$ is placed at $t_P=\min(0,t)-\Delta T$. 
The two- and three-point functions share the same wall source, so the same energy gaps govern the suppression of excited states propagating from it. We therefore use the effective-mass plateau of the wall-source--point-sink two-point correlator $\langle\phi_P(t)\phi_P^\dagger(0)\rangle$ to guide our choice of $\Delta T$. This procedure assumes similar excited-state suppression in the three-point functions. In lattice units, the effective mass $m_{\rm eff}(t)$ is defined by
\begin{equation}
\frac{\langle\phi_P(t+1)\phi_P^\dagger(0)\rangle}{\langle\phi_P(t)\phi_P^\dagger(0)\rangle}
=\frac{\cosh\!\left[m_{\rm eff}(t)(t+1-T/2)\right]}{\cosh\!\left[m_{\rm eff}(t)(t-T/2)\right]}.
\label{eq:source_meff}
\end{equation}
Based on the pion and kaon plateaus in Fig.~\ref{fig:source_meff}, we choose $\Delta T=12a$ on 48I and $18a$ on 64I, approximately $1.4$ and $1.5\,\mathrm{fm}$, respectively. The values of $\Delta T$ used for all ensembles are listed in Table~\ref{table:ens}.

\begin{figure*}
\centering
\includegraphics[width=.7\textwidth]{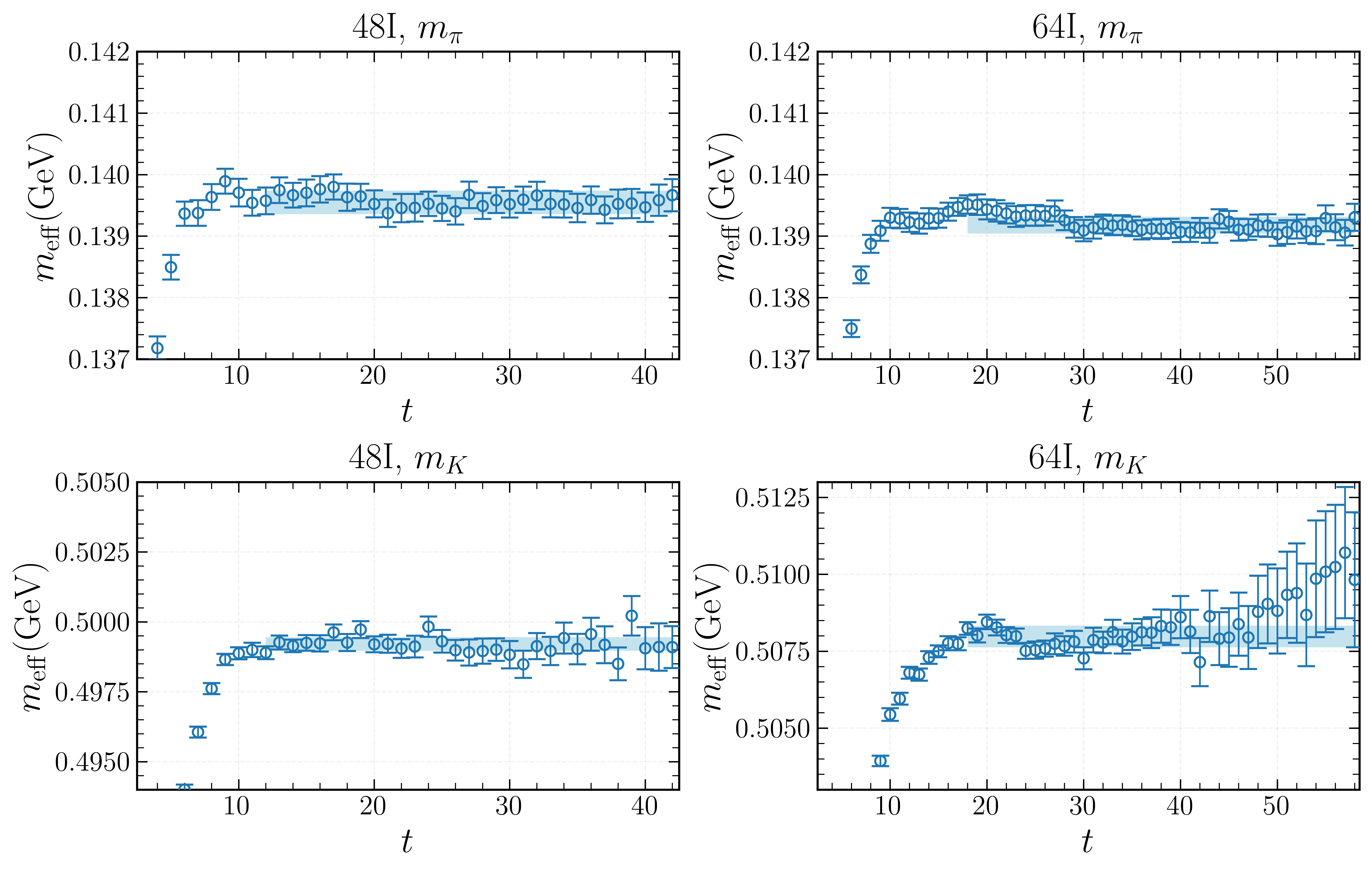}
\captionsetup{labelfont={color=revision},textfont={color=revision}}
\caption{Pion and kaon effective masses from wall-source--point-sink correlators on 48I and 64I, reproduced from Ref.~\cite{Boyle:2025uuh}. Blue bands show the meson masses and statistical errors from fits over $[\Delta T,T-\Delta T]$. The horizontal axis is $t$ in lattice units; masses are in GeV.}
\label{fig:source_meff}
\end{figure*}

\endgroup

\section{Lattice ensembles, \textcolor{revision}{parameters,} and complete results for all schemes}
\label{app:ensemble_results}
Table~\ref{table:ens} summarizes the parameters of the lattice ensembles used in this work: the $N_f=2+1$ physical-pion-mass domain-wall ensembles 24D, 32D, 48I, and 64I, together with the partially quenched setups 64I-pq1 and 64I-pq2 employed for the kaon-mass correction (Sec.~\ref{app:meson_mass_correction}).

\begin{table*}
	\centering\small\setlength{\tabcolsep}{3pt}
	\begin{tabular}{cccccccccc}
		\hline\hline
		Ensemble & $a^{-1}$[GeV] & $L^3\times T$ &$a\,m_l$ &$a\,m_s$  & $m_\pi$/MeV & $m_K$/MeV & $N_{\text{conf}}$ &$m_\pi L$  & \textcolor{revision}{$\Delta T/a$} \\
		\hline
    $24$D & $1.023$ & $24^3\times 64$ & $0.00107$ & $0.0850$ & $142.72(23)$ & $515.36(27)$ & $145$ & $3.3$ & \textcolor{revision}{$8$} \\
    $32$D & $1.023$ & $32^3\times 64$ & $0.00107$ & $0.0850$ & $142.51(22)$ & $515.39(25)$ & $63$ & $4.5$ & \textcolor{revision}{$8$} \\
    $48$I & $1.730$ & $48^3\times 96$ & $0.00078$ & $0.0362$ & $139.55(19)$ & $499.21(24)$ & $112$ & $3.9$ & \textcolor{revision}{$12$} \\
    $64$I & $2.359$ & $64^3\times 128$ & $0.000678$ & $0.02661$ & $139.18(14)$ & $507.98(35)$ & $119$ & $3.8$ & \textcolor{revision}{$18$} \\
    $64$I-pq1 & $2.359$ & $64^3\times 128$ & $0.0006203$ & $0.02539$ & $135.14(19)$ & $496.50(81)$ & $31$ & $3.7$ & \textcolor{revision}{$18$} \\
    $64$I-pq2 & $2.359$ & $64^3\times 128$ & $0.000678$ & $0.02539$ & $139.13(19)$ & $497.01(80)$ & $31$ & $3.8$ & \textcolor{revision}{$18$} \\
		\hline\hline
	\end{tabular}
	\caption{\label{table:ens}Parameters of the lattice ensembles used in this work: inverse lattice spacing $a^{-1}$ (GeV), $L^3\times T$, quark masses, $\pi$/$K$ masses, number of configurations $N_{\text{conf}}$, $m_\pi L$, \textcolor{revision}{and the source--current separation $\Delta T$. The quark masses are valence masses.}}
\end{table*}

\begingroup
\color{revision}
\hypersetup{linkcolor=revision,citecolor=revision,urlcolor=revision}
Table~\ref{tab:analysis_inputs} lists the common physical inputs. The integration kernels, point-particle subtraction terms, and finite-volume and ATW models use the meson masses measured on each ensemble. The physical meson masses enter the tree-level ratio $R_{e/\mu}^{(0)}$, real-photon phase space, and correction of the kaon results to the physical point. 

\par\medskip
\noindent\begin{minipage}{\columnwidth}
\color{revision}
\captionsetup{labelfont={color=revision},textfont={color=revision}}
\centering\small
\captionof{table}{Common physical inputs. Masses are in MeV. The electromagnetic coupling and physical masses are taken from the PDG at the quoted precision~\cite{ParticleDataGroup:2024cfk}.}\label{tab:analysis_inputs}
\begin{tabular}{ll}\toprule
Input & Value\\\midrule
$\alpha^{-1}$ & $137.036$\\
$(m_e,m_\mu)$ & $(0.51099895,105.6583755)$\\
$(m_{\pi^\pm},m_{K^\pm})$ & $(139.57039,493.677)$\\\bottomrule
\end{tabular}
\end{minipage}
\par\medskip
\endgroup

For completeness, we list the results for all methods, both on each ensemble and after the continuum extrapolation. Table~\ref{tab:M2full} gives the per-ensemble results for $\delta_{\text{SD}}$ before continuum extrapolation and mass correction, including the four virtual-correction schemes, Feynman/Coulomb gauge $\times$ with/without the $R_{f_P}$ technique, as well as the real-photon-emission result. \textcolor{revision}{All results include the ATW treatment and finite-volume corrections used in the analysis.}

Extrapolating the 48I result together with the mass-corrected 64I result to the continuum limit yields the results shown in Table~\ref{tab:M2}. In the main text, we take the Coulomb-gauge results with the $R_{f_P}$ technique as our final values. 

\begin{table*}
    \centering\small\setlength{\tabcolsep}{4pt}
    \begin{tabular}{l ccccc}
    \hline\hline
    $\pi$ & \multicolumn{4}{c}{$\delta_{\text{SD}}^{\text{vir}}/10^{-3}$} & \multirow{2}{*}{$\delta_{\text{SD}}^{\text{real}}/10^{-3}$} \\
    Ensemble & Feyn.(no $R_{f_P}$) & Feyn.(w. $R_{f_P}$) & Coul.(no $R_{f_P}$) & Coul.(w. $R_{f_P}$) & \\
    \hline
    24D & \textcolor{revision}{$0.645(96)$} & \textcolor{revision}{$0.562(19)$} & \textcolor{revision}{$0.711(53)$} & \textcolor{revision}{$0.658(9)$} & \textcolor{revision}{$0.733(33)$} \\
    32D & \textcolor{revision}{$0.351(94)$} & \textcolor{revision}{$0.577(22)$} & \textcolor{revision}{$0.513(53)$} & \textcolor{revision}{$0.658(9)$} & \textcolor{revision}{$0.737(30)$} \\
    48I & \textcolor{revision}{$0.803(74)$} & \textcolor{revision}{$0.601(20)$} & \textcolor{revision}{$0.706(42)$} & \textcolor{revision}{$0.576(9)$} & \textcolor{revision}{$0.652(36)$} \\
    64I & \textcolor{revision}{$0.719(76)$} & \textcolor{revision}{$0.585(24)$} & \textcolor{revision}{$0.664(43)$} & \textcolor{revision}{$0.577(10)$} & \textcolor{revision}{$0.636(37)$} \\
    64I-pq1 & \textcolor{revision}{$0.739(154)$} & \textcolor{revision}{$0.617(40)$} & \textcolor{revision}{$0.669(90)$} & \textcolor{revision}{$0.592(22)$} & \textcolor{revision}{$0.509(52)$} \\
    64I-pq2 & \textcolor{revision}{$0.758(149)$} & \textcolor{revision}{$0.628(38)$} & \textcolor{revision}{$0.674(87)$} & \textcolor{revision}{$0.592(21)$} & \textcolor{revision}{$0.562(49)$} \\
    \hline\hline
    $K$ & \multicolumn{4}{c}{$\delta_{\text{SD}}^{\text{vir}}/10^{-3}$} & \multirow{2}{*}{$\delta_{\text{SD}}^{\text{real}}$ (not in $R_{e/\mu}$)} \\
    Ensemble & Feyn.(no $R_{f_P}$) & Feyn.(w. $R_{f_P}$) & Coul.(no $R_{f_P}$) & Coul.(w. $R_{f_P}$) & \\
    \hline
    24D & \textcolor{revision}{$1.435(98)$} & \textcolor{revision}{$1.708(89)$} & \textcolor{revision}{$1.519(40)$} & \textcolor{revision}{$1.665(44)$} & \textcolor{revision}{$1.598(30)$} \\
    32D & \textcolor{revision}{$1.244(105)$} & \textcolor{revision}{$1.870(87)$} & \textcolor{revision}{$1.384(37)$} & \textcolor{revision}{$1.718(45)$} & \textcolor{revision}{$1.514(30)$} \\
    48I & \textcolor{revision}{$1.297(89)$} & \textcolor{revision}{$1.381(65)$} & \textcolor{revision}{$1.393(34)$} & \textcolor{revision}{$1.439(34)$} & \textcolor{revision}{$1.279(25)$} \\
    64I & \textcolor{revision}{$1.283(88)$} & \textcolor{revision}{$1.343(79)$} & \textcolor{revision}{$1.425(32)$} & \textcolor{revision}{$1.457(40)$} & \textcolor{revision}{$1.358(28)$} \\
    64I-pq1 & \textcolor{revision}{$1.349(205)$} & \textcolor{revision}{$1.297(166)$} & \textcolor{revision}{$1.441(60)$} & \textcolor{revision}{$1.414(79)$} & \textcolor{revision}{$1.217(45)$} \\
    64I-pq2 & \textcolor{revision}{$1.344(201)$} & \textcolor{revision}{$1.325(159)$} & \textcolor{revision}{$1.439(60)$} & \textcolor{revision}{$1.427(78)$} & \textcolor{revision}{$1.214(43)$} \\
    \hline\hline
    \end{tabular}
    \caption{\label{tab:M2full}Per-ensemble results for $\delta_{\text{SD}}$ before the continuum extrapolation and kaon-mass correction. The errors in parentheses include only the statistical errors. }
\end{table*}

\begin{table*}
    \centering\small\setlength{\tabcolsep}{4pt}
    \begin{tabular}{l cc}
    \hline\hline
    Source & pion $\delta^{\text{vir}}_{\text{SD}}/10^{-3}$ & kaon $\delta^{\text{vir}}_{\text{SD}}/10^{-3}$ \\
    \hline
    Lat. Feyn. (no $R_{f_P}$) & \textcolor{revision}{$0.621(187)_{\mathrm{stat}}(295)_{\mathrm{FV}}(2)_{a^4}(9)_{\mathrm{qdisc}}$} & \textcolor{revision}{$1.205(245)_{\mathrm{stat}}(192)_{\mathrm{FV}}(5)_{a^4}(17)_{\mathrm{qdisc}}$} \\
    Lat. Feyn. (w. $R_{f_P}$) & \textcolor{revision}{$0.567(55)_{\mathrm{stat}}(14)_{\mathrm{FV}}(2)_{a^4}(8)_{\mathrm{qdisc}}$} & \textcolor{revision}{$1.185(214)_{\mathrm{stat}}(162)_{\mathrm{FV}}(4)_{a^4}(17)_{\mathrm{qdisc}}$} \\
    Lat. Coul. (no $R_{f_P}$) & \textcolor{revision}{$0.616(104)_{\mathrm{stat}}(197)_{\mathrm{FV}}(2)_{a^4}(9)_{\mathrm{qdisc}}$} & \textcolor{revision}{$1.372(86)_{\mathrm{stat}}(135)_{\mathrm{FV}}(5)_{a^4}(19)_{\mathrm{qdisc}}$} \\
    Lat. Coul. (w. $R_{f_P}$) & \textcolor{revision}{$0.579(24)_{\mathrm{stat}}(0)_{\mathrm{FV}}(2)_{a^4}(8)_{\mathrm{qdisc}}$} & \textcolor{revision}{$1.359(101)_{\mathrm{stat}}(53)_{\mathrm{FV}}(5)_{a^4}(19)_{\mathrm{qdisc}}$} \\
    ChPT & $0.530(110)$ & $1.350(110)$ \\
    \hline
    \hline
    Source & pion $\delta^{\text{real}}_{\text{SD}}/10^{-3}$ & kaon $\delta^{\text{real}}_{\text{SD}}$ (not in $R_{e/\mu}$) \\
    \hline
    Lat. & \textcolor{revision}{$0.617(88)_{\mathrm{stat}}(3)_{\mathrm{FV}}(2)_{a^4}(9)_{\mathrm{qdisc}}$} & \textcolor{revision}{$1.215(70)_{\mathrm{stat}}(84)_{\mathrm{FV}}(5)_{a^4}(17)_{\mathrm{qdisc}}$} \\
    ChPT & $0.730$ & -- \\
    \hline\hline
    \end{tabular}
    \caption{\label{tab:M2}Continuum-extrapolated results for the SD corrections $\delta_{\text{SD}}$ for $\pi$ and $K$, compared with ChPT predictions. The kaon results are corrected to the physical $K^\pm$ mass; see Sec.~\ref{app:meson_mass_correction}. Upper part: the SD correction $\delta_{\text{SD}}^{\text{vir}}$ for the four schemes, Feynman/Coulomb gauge $\times$ with/without the $R_{f_P}$ technique. Lower part: the SD correction $\delta_{\text{SD}}^{\text{real}}$. \textcolor{revision}{The uncertainty labels are the same as in Table~\ref{tab:SD_final}.}}
\end{table*}

\clearpage
\end{document}